\documentclass[twocolumn]{article}
\usepackage[utf8]{inputenc}
\usepackage{graphicx}
\usepackage{amsmath,amssymb,amsthm}
\usepackage{algorithm,algpseudocode}
\usepackage{url}
\usepackage{hyperref}
\usepackage[top=1in, bottom=1in, left=0.5in, right=0.5in]{geometry}
\usepackage{abstract}
\usepackage{authblk}
\usepackage{float}
\usepackage{subfig}
\usepackage{color}

\title{Local equations describe unreasonably efficient stochastic algorithms in random K-SAT}
\author[*, 1,2,3]{David Machado}
\author[1]{Jonathan González-García}
\author[1]{Roberto Mulet}
\affil[1]{Center for Complex Systems and Department of Theoretical Physics, University of Havana, Cuba}
\affil[2]{Dipartimento di Fisica, Sapienza Università di Roma, Italy}
\affil[3]{CNR - Nanotec, unità di Roma, Italy}
\affil[*]{Corresponding author: \href{mailto:david.machado@uniroma1.it}{\color{blue} david.machado@uniroma1.it}}

\begin{document}

\twocolumn[
\maketitle

\begin{abstract}
Despite significant advances in characterizing the highly nonconvex landscapes of constraint satisfaction problems, the good performance of certain algorithms in solving hard combinatorial optimization tasks remains poorly understood. This gap in understanding stems largely from the lack of theoretical tools for analyzing their out-of-equilibrium dynamics. To address this challenge, we develop a system of approximate master equations that capture the behavior of local search algorithms in constraint satisfaction problems. Our framework shows excellent qualitative agreement with the phase diagrams of two paradigmatic algorithms: Focused Metropolis Search (FMS) and greedy-WalkSAT (G-WalkSAT) for random 3-SAT. The equations not only confirm the numerical observation that G-WalkSAT's algorithmic threshold is nearly parameter-independent, but also successfully predict FMS's threshold beyond the clustering transition. We also exploit these equations in a decimation scheme, demonstrating that the computed marginals encode valuable information about the local structure of the solution space explored by stochastic algorithms. Notably, our decimation approach achieves a threshold that surpasses the clustering transition, outperforming conventional methods like Belief Propagation-guided decimation. These results challenge the prevailing assumption that long-range correlations are always necessary to describe efficient local search dynamics and open a new path to designing efficient algorithms to solve combinatorial optimization problems.
\end{abstract}
]

The study of satisfiability problems is a critical area at the frontier between computational complexity and statistical physics. In particular, the random K-SAT problem has served as a benchmark for studying algorithmic performance and for gaining fundamental insight into phase transitions and the dynamical behavior of complex systems. The problem is defined through $N$ Boolean variables and $M$ clauses, each clause being a disjunction of $K$ distinct literals (variables or their negations) chosen uniformly at random. It asks whether there exists a truth assignment that satisfies all clauses for a given value of the ratio $\alpha= M/N$. 
Varying this ratio $\alpha$, the system exhibits two thresholds. The onset of satisfiability, or SAT-UNSAT transition, occurs at approximately $\alpha_s \approx 4.267$ \cite{mezardksat2002, MertensKSAT2006}, but before there is also a dynamical transition at $\alpha_d \approx 3.86$ \cite{mezardksat2002, KrzakalaKSAT2007, Montanari_2008} where solutions cluster into distinct regions. 

The identification of the latter has been fundamental in understanding the nature of combinatorial optimization problems and their computational complexity. However, contradictions and open questions persist, particularly concerning algorithmic hardness and the alignment of theoretical predictions with empirical observations\cite{SemerjianwalkSAT2003, Semerjiandyn2004, ArdeliusASAT2006, AlavaPNAS2008, Lage2014message,LemoyVFMS2015,Monasson2004,Achlioptas08,Budzynski_2019}. One of the central challenges here is reconciling the predictions of equilibrium statistical mechanics with the behavior of practical algorithms.  For example, empirical studies of algorithms like WalkSAT \cite{AurellNIPS2004WalkSAT, Seitz_2005} reveal that it starts to fail well beyond the clustering threshold. Furthermore, if the parameters are tuned correctly, the Focused Metropolis Search (FMS)\cite{Seitz_2005,AlavaPNAS2008} finds solutions beyond $\alpha_d$ and very close to the SAT-UNSAT transition. 

Using the cavity method in depth, it was possible to deduce other thresholds such as the condensation transition $\alpha_c$ \cite{KrzakalaKSAT2007} and the rigidity threshold \cite{achlioptas2006solution, ZdeborovaPRE2007qcol}.  A stronger condition for the onset of hardness, called the Overlap Gap Property, has been recently developed in Refs. \cite{gamarnik2017limits, gamarnik2018finding, gamarnik2021overlap}. But none of these thresholds seems relevant to understanding algorithm dynamics in random K-SAT when $K$ is small. Moreover, the authors in \cite{BaldassiJSTAT2016,BaldassiPNAS2016,MariaChiara2023PRX} have shown that exploiting non-equilibrium measures could lead to {\it unreasonable effective algorithms}, under the hypothesis that their out-of-equilibrium dynamics focuses on particularly entropic regions in the space of solutions. These discrepancies between the results of equilibrium statistical mechanics and algorithms underscore the need for a non-equilibrium approach to study local search algorithms for combinatorial optimization problems.

Another area of active debate concerns the role of algorithmic parameters and their impact on performance.  For example, the behavior of WalkSAT with greedy steps \cite{BarthelwalkSAT2003} (G-WalkSAT) demonstrates pathological stagnation when all steps are greedy. In this case, the algorithm becomes trapped at low energies without finding solutions (see the appendixes for numerical confirmation and discussion). Additionally, the interplay between greedy and stochastic strategies in solvers raises fundamental questions about the trade-offs between exploration and exploitation in complex landscapes. 

In this work, we approximate the master equation for local search algorithms by adapting the Conditional Dynamic Approximation (CDA) \cite{CDA-Pspin} to single instances. We study the dynamics of two prominent stochastic local search algorithms, G-WalkSAT \cite{BarthelwalkSAT2003} and Focused Metropolis Search (FMS) \cite{Seitz_2005}, in solving random 3-SAT instances. Our analysis leverages both numerical simulations and a theoretical framework, comparing its predictions with two techniques in the literature, the Cavity Master Equation (CME) \cite{CME-Pspin,CME-PRL,hCME} and the Dynamic Independent-Neighbors Approximation (DINA) \cite{BarthelwalkSAT2003, Semerjiandyn2004}.

Through CDA, we can predict the behavior of these algorithms outperforming all previous approaches across varying constraint densities and algorithmic parameters. The dynamics reveal different regimes depending on the parameters $q$ (for G-WalkSAT) and $\eta$ (for FMS). We present comprehensive phase diagrams that delineate the limits of algorithmic success, marking the transitions from polynomial to non-polynomial time solutions. We also extend our findings beyond the theoretical description of individual algorithms by incorporating a decimation strategy. Our CDA-guided decimation finds solutions in the supposedly hard region $\alpha_d < \alpha < \alpha_s$. This offers new avenues for enhancing solver efficiency and studying the local statistics of the solutions typically found by a specific algorithm. %These results underscore the importance of integrating advanced theoretical tools with empirical observations to uncover the mechanisms governing algorithmic success and failure in hard combinatorial problems.

%Our results underscore the importance of integrating advanced theoretical tools with empirical observations to uncover the mechanisms governing algorithmic success and failure in hard combinatorial problems. They also bring a deeper understanding of the algorithmic performance in combinatorial optimization problems, bridging gaps between computational experiments and theoretical predictions. Moreover, they pave the way for designing more robust solvers tailored to the structural challenges posed by constraint satisfaction problems. 

\section{Conditional Dynamic Approximation}

In this section, we introduce the set of equations that describe the algorithmic dynamics. We first do it in a general setting and then apply them to our algorithms of interest. These equations are written for problems with $N$ discrete variables (denoted by the vector $\vec{\sigma}=\{\sigma_1, \ldots, \sigma_N \}$) on sparse graphs and asynchronous local dynamics. Although similar equations were already used by us in Ref. \cite{CDA-Pspin}, their use was restricted to the time evolution of average quantities in the K-XORSAT problem on random regular graphs. Here, we show a more general version that could help the reader to extend the results to other hard combinatorial optimization problems with random connectivity and/or other types of local disorder. To prove their relevance, we apply them to single instances of the random K-SAT problem in graphs with many variables, Poisson connectivity, and disordered interactions.

%and apply them to single instances of the random K-SAT problem in graphs with many variables and random connectivity. We will present our equations in a form that will help the reader to extend the results to other hard combinatorial optimization problems like hypergraph bicoloring \cite{} or the same K-XORSAT \cite{} in Poissonian graphs. For simplicity, we will stick to binary variables $\sigma_i=\pm1$, but it is straightforward to write everything for general $q$-state variables, necessary for other problems like $q$-coloring \cite{}.{\bf RW}

In general, an algorithm consists of a set of rules to transform a candidate solution $\vec{\sigma}$ to another $\vec{\sigma}'$. We assume that this information can be encoded in a function $r_i(\sigma_i, \vec{\sigma}_{\partial i})$. This is the rate at which the $i$-th variable will adopt the value $-\sigma_i$, given its current value $\sigma_i$ and the configuration of its neighbors $\vec{\sigma}_{\partial i}$. 

%Let us assume we choose a specific algorithm to solve our problem. In general, the definition of the algorithm would consist of a set of rules that explain how to transform a candidate solution $\vec{\sigma}$ to another one $\vec{\sigma}'$. We assume that this information can be encoded into a function $r_i(\sigma_i, \vec{\sigma}_{\partial i})$. This is the probability that the $i$-th variable will adopt the value $-\sigma_i$, given its current value $\sigma_i$ and the configuration of its neighbors $\vec{\sigma}_{\partial i}$. Taking into account these local transition rules $r_i$ is not very restrictive. Indeed, it is possible to write suitable $r_i$ functions for relevant examples like traditional Monte Carlo Markov Chains (MCMC), Focused Metropolis Search (FMS) or walksat in its several variants. 

%To describe the dynamics, we will use differential equations in a continuous time formulation. If the process is sufficiently \textit{smooth}, one should be able to write time derivatives of the probability $P_{t}(\vec{\sigma})$ to find the system in a given configuration at time $t$. The auxiliary continuous time process would be sufficiently close to the real discrete time chain. This approach is not novel since one can find several other uses in the literature \cite{}, and can be even made rigorous \cite{}.

Within this scenario, the more general equation to use in continuous time is the well-known master equation \cite{vanKampen92} for the dynamics:

\begin{eqnarray}
\frac{dP_t(\vec{\sigma})}{dt} = -\sum_{i=1}^{N} r_i(\sigma_i, \vec{\sigma}_{\partial i}) \, P_t(\vec{\sigma}) \nonumber \\
+ \sum_{i=1}^{N}  r_i(-\sigma_i, \vec{\sigma}_{\partial i}) \, P_t(F_{i}[\vec{\sigma}]) 
 \label{eq:mas_eq}
\end{eqnarray}
where $F_{i}[\vec{\sigma}]$ is the vector  $\{\sigma_1, \ldots, \sigma_{i-1}, -\sigma_i, \sigma_{i+1}, \ldots, \sigma_N\}$, obtained after substituting $\sigma_i$ by $-\sigma_i$ in $\vec{\sigma}$.

The probability distribution $P_t(\vec{\sigma})$ occupies a high-dimensional space that is computationally intractable to explore exhaustively, making stochastic sampling essentially mandatory. However, we are often primarily interested in computing local observables. For instance, in the random K-SAT problem with $N$ variables and $M$ clauses, each satisfied clause contributes zero to the system's energy while unsatisfied clauses contribute one. A key question is how the system's energy evolves under specific stochastic dynamics.

%the whole distribution $P_{t}(\vec{\sigma})$, and we settle if we can predict the time evolution of some relevant magnitude. In constraint satisfaction problems, like random K-SAT, we have $M$ functions $\varphi(\vec{\sigma}_a)$ to be simultaneously satisfied ($\varphi$ evaluates to '$0$' when satisfied and to '$1$' otherwise). Here, the index $a$ goes from one to $M$, and $\vec{\sigma}_a$ is the configuration of the variables involved in the $a$-th constraint. In the particular case of the random K-SAT problem, those constraints are the $M$ clauses of the Boolean formula to be satisfied.

To compute the evolution of the joint probability $P_t(\vec{\sigma}_a)$ associated with a specific clause, we can marginalize Eq. \ref{eq:mas_eq} over all vectors $\vec{\sigma}$ that share the same local $\vec{\sigma}_a$. The result is:

\begin{eqnarray}
\frac{dP_t(\vec{\sigma}_a)}{dt} = -\sum_{i \in a} \sum_{\vec{\sigma}_{\partial i \setminus a}} r_i(\sigma_i, \vec{\sigma}_{\partial i}) \, P_t(\sigma_i, \vec{\sigma}_{\partial i}) \nonumber \\
+ \sum_{i\in a} \sum_{\vec{\sigma}_{\partial i \setminus a}} r_i(-\sigma_i, \vec{\sigma}_{\partial i}) \, P_t(-\sigma_i, \vec{\sigma}_{\partial i}) 
 \label{eq:mas_eq_sa}
\end{eqnarray}

\begin{figure}[ht]
\centering
\includegraphics[width=.3\linewidth]{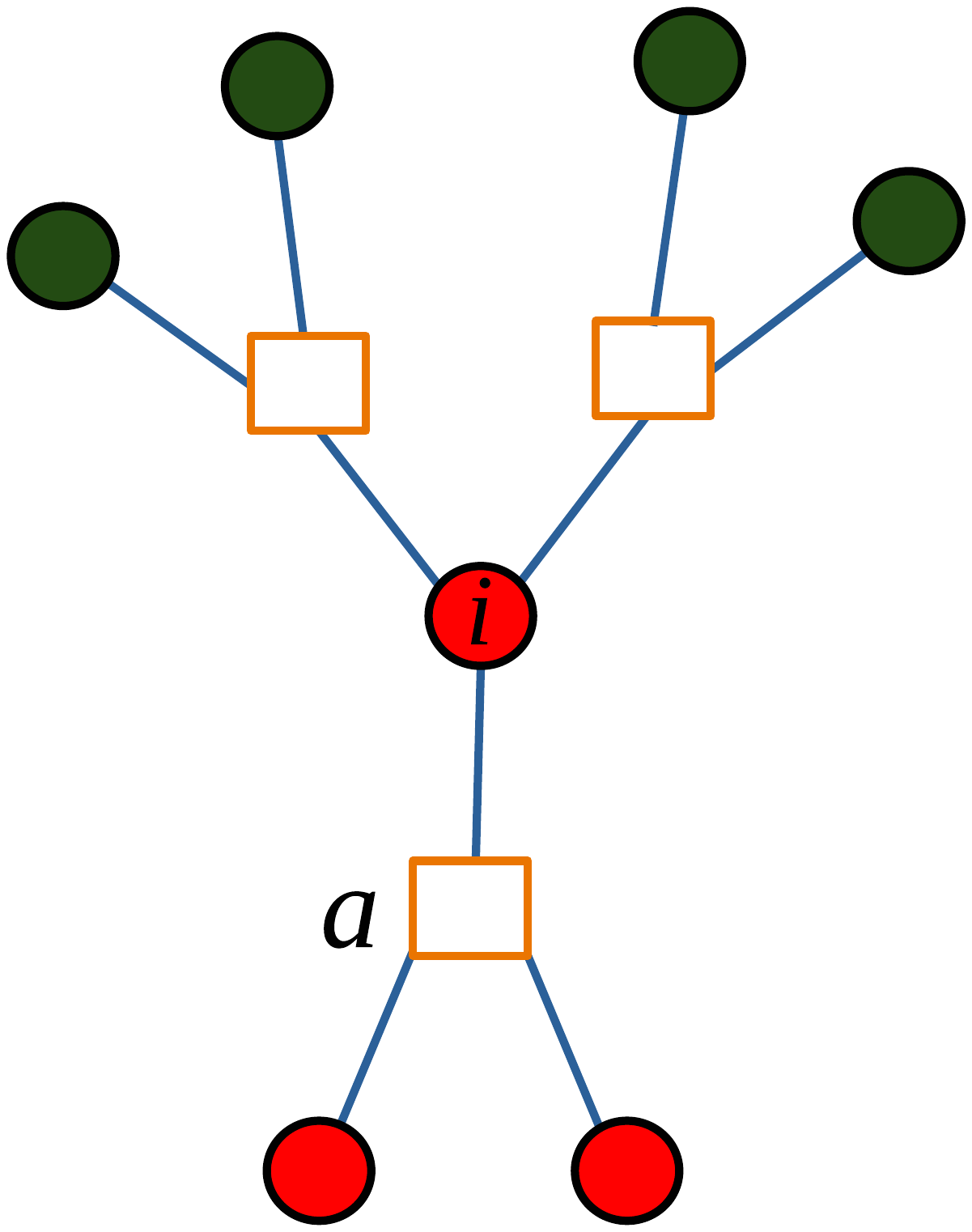}
\includegraphics[width=.3\linewidth]{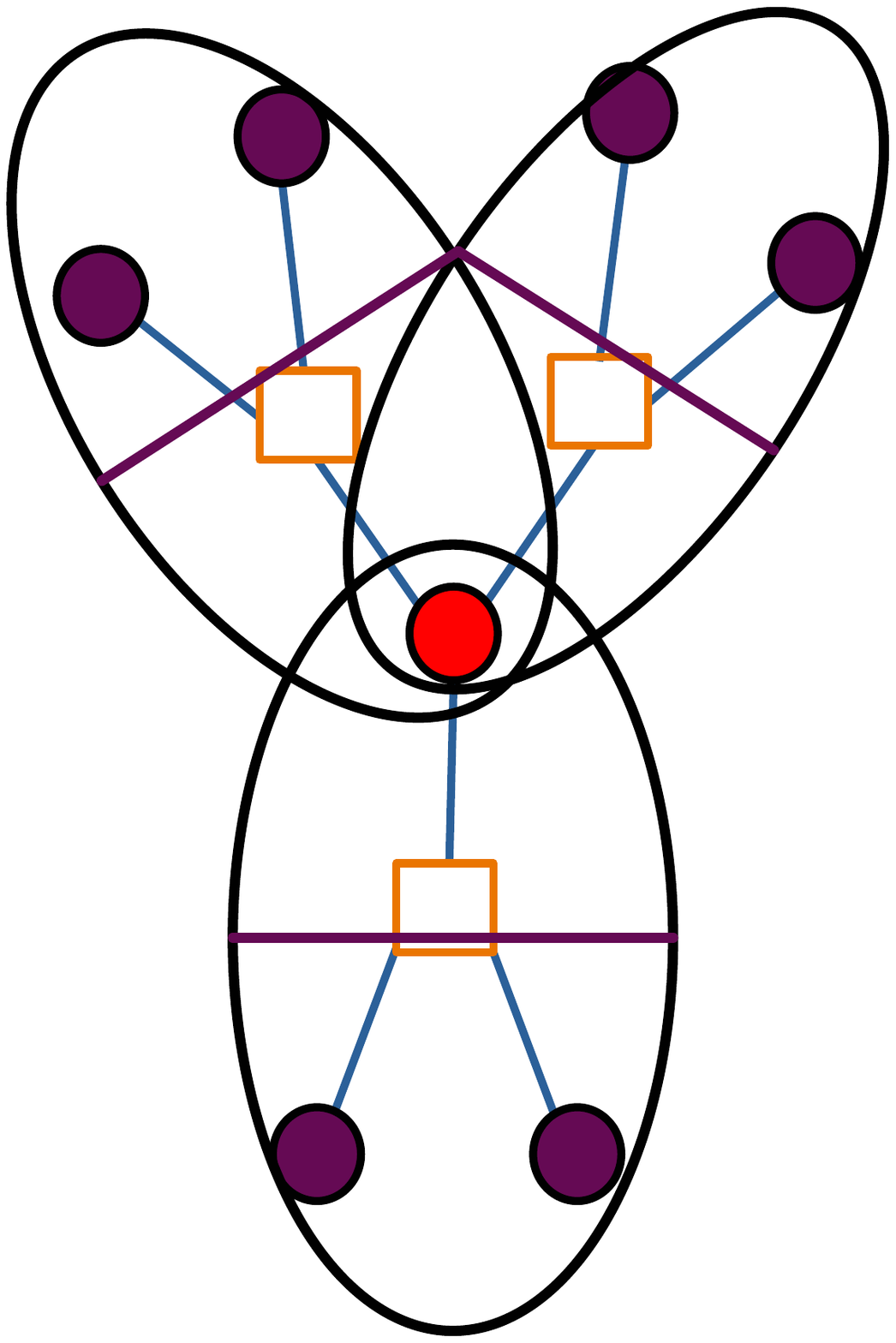}
    \caption{Portion of the factor graph involved in Eqs. \ref{eq:mas_eq_sa}, \ref{eq:factorization} and \ref{eq:CDA1}. Variable nodes are represented with circles and are connected to a second type of node symbolizing the constraints (squares in the figure). ({\bf left panel})The nodes participating in the $a$-th clause are colored in red. The other neighbors of the $i$-th variable are colored in green. ({\bf right panel}) The factors in the first line of Eq. \ref{eq:factorization}, each associated with a different conditional probability, are surrounded by black ovals.}
    \label{fig:illustration_CDA}
\end{figure}

Fig. \ref{fig:illustration_CDA} illustrates the portion of the graph where Eq. \ref{eq:mas_eq_sa} is operating. We use factor graphs, aware that the right notation for the variables $i$ connected to the node $a$ would be $\partial a$, and the right relation would be $i \in \partial a$. However, we chose a simpler notation to keep our equations compact. In Eq. \ref{eq:mas_eq_sa}, the sum $\sum_{\vec{\sigma}_{\partial i \setminus a}}$ is taken over all the configurations of the neighbors of $\sigma_i$ that are not involved in the $a$-th clause (colored green in the figure). Instead, the sum $\sum_{i \in a}$ is over all the indexes of the variables $\sigma_i$ involved in the $a$-th clause (colored red). 

The Eq. \ref{eq:mas_eq_sa} is not closed and, as presented here, is the first step in finding a proper closure for the master equation. To do so, we assume that the dynamics are well described by local correlations and use one of the factorizations proposed in Ref. \cite{CDA-Pspin}:

%The solution of Eq. \ref{eq:mas_eq_sa} would give $P_t(\vec{\sigma}_a)$ at all times if we were also able to compute $P_t(\sigma_i, \vec{\sigma}_{\partial i})$ for all $i \in a$. However, if we write the differential equation for the latter, we find new probabilities defined over larger portions of the graph. Iterating, one gets a hierarchical system of equations that does not stop until we again get Eq. \ref{eq:mas_eq}. We would have gained nothing. This is why we say that Eq. \ref{eq:mas_eq_sa}, as presented, is not closed. Precisely at this point, we will provide an approximate {\bf {\textit{dynamic closure}}}.

%To do so, we assume that the dynamics are well described by local correlations and use one of the factorizations proposed in Ref. \cite{CDA-Pspin}:

\begin{eqnarray}
P_t(\sigma_i, \vec{\sigma}_{\partial i}) \approx P_t(\sigma_i)  \prod_{b \in \partial i} P_t(\vec{\sigma}_{b \setminus i} \mid \sigma_i) 
 \label{eq:factorization}
\end{eqnarray}

Eq. \ref{eq:factorization} assumes that the neighbors of the $i$-th variable are independent once conditioned on $\sigma_i$. Recalling the definition of conditional probability $P_t(\vec{\sigma}_{b \setminus i} \mid \sigma_i) = P_t(\vec{\sigma}_b)/P_t(\sigma_i)$ we get a solvable approximate master equation:

\begin{eqnarray}
\frac{dP_t(\vec{\sigma}_a)}{dt} = -\sum_{i \in a} \sum_{\vec{\sigma}_{\partial i \setminus a}} r_i(\sigma_i, \vec{\sigma}_{\partial i}) \, \Big[ \prod_{b \in \partial i \setminus a} \frac{P_t(\vec{\sigma}_{b})}{P_t(\sigma_i)} \Big] P_t(\vec{\sigma}_a) \nonumber \\
+  \sum_{i\in a} \sum_{\vec{\sigma}_{\partial i \setminus a}} r_i(-\sigma_i, \vec{\sigma}_{\partial i}) \, \Big[ \prod_{b \in \partial i \setminus a} \frac{P_t(F_{i}[\vec{\sigma}_{b}])}{P_t(-\sigma_i)} \Big] P_t(F_{i}[\vec{\sigma}_a])
 \label{eq:CDA1}
\end{eqnarray}

This is what we call Conditional Dynamic Approximation (CDA). In the simpler case with pairwise interactions {\em and dynamic rules } $r_i$ decomposable as a sum over neighbors ($r_i(\sigma_i, \sigma_{\partial i}) = \sum_{k \in \partial i} r_i(\sigma_i, \sigma_k)$), the CDA reduces to the well-known Pair-Based Mean Field approximation \cite{cator2012second, mata2013pair}. The latter has been particularly useful in the study of epidemic spreading in networks \cite{cator2012second, mata2013pair, pastor2015epidemic, silva19PQMF, silva20PQMF}.

When the dynamics satisfy detailed balance, the fixed point of the CDA equations coincides with the solution obtained with the equilibrium cavity method\cite{CME-AvAndBP}. However, far from equilibrium, the precise form of the non-equilibrium probability distribution remains unknown. The accuracy of CDA hinges on the assumption that local correlations govern the temporal evolution. Consequently, the predictive success of this approximation for a given algorithm can serve to diagnose the dominant dynamical mechanisms at play. We expect that incorporating higher-order cluster approximations (e.g., \cite{hCME,CDA-Pspin}) — which account for extended spatial or temporal correlations — could provide systematic improvements.

%these equations is consistent with the solution of the equilibrium cavity method \cite{CME-AvAndBP}. Out of equilibrium, there are no exact results on the shape of the probability distribution. We expect the approximation to be accurate if local correlations are enough to describe the temporal evolution of the system. Therefore, the accuracy in the predictions for a specific algorithm can help assess the key mechanism involved in its dynamics. However, we expect approximations considering larger clusters of variables (see for example \cite{hCME,CDA-Pspin}) to improve the results.

\section{Two algorithms}

FMS and G-WalkSAT are local search algorithms that do not fulfill the detailed balance condition. Given a configuration $\vec{\sigma}$ of the variables in a random K-SAT formula with $M$ clauses, they only propose changes in the variables that belong to one of the unsatisfied clauses. To apply the CDA, we need to properly define the corresponding rules $r_i(\sigma_i, \vec{\sigma}_{\partial i})$.

%FMS and G-WalkSAT are local search algorithms that do not fulfill the detailed balance condition. Given a configuration $\vec{\sigma}$ of the variables in a random K-SAT formula with $M$ clauses, they will only propose changes in the variables that belong to one of the unsatisfied clauses. Fortunately, that is not a problem for our equations. To apply the CDA, we only need to define the corresponding rules $r_i(\sigma_i, \vec{\sigma}_{\partial i})$.

{\bf FMS} selects a candidate variable to flip as follows: first, it takes an {\it unsatisfied} clause uniformly at random; then, it takes a variable inside that clause, also uniformly at random. Once the index $i$ of the variable is known, it will propose the change $\sigma_i \to -\sigma_i$ with a probability analogous to the Metropolis rule \cite{Metropolis1953}, well-known in the field of Monte Carlo Markov Chains. Thus, the probability of flipping the $i$-th variable is:

\begin{equation}
 r_i^{\text{FMS}}(\sigma_i, \vec{\sigma}_{\partial i}) = \frac{E_i}{K E} \, \text{min}\{1, \eta^{\Delta E} \} \label{eq:rate_FMS}
\end{equation}
where $E_i$ is the number of unsatisfied clauses that contain the $i$-th variable, $E$ is the total number of unsatisfied clauses, and $\Delta E$ is the change in $E$ produced by flipping the variable. The algorithmic parameter $\eta \in [0, 1]$ controls the noise in the dynamics. When $\eta=1$, all the proposed changes are accepted, and one recovers the famous random WalkSAT algorithm \cite{PapadimitrouWalkSAT}. When $\eta=0$, the algorithm becomes greedy and only accepts changes that diminish the number of unsatisfied clauses (with $\Delta E < 0$).

The reader should note that the number of unsatisfied clauses $E$ depends on the whole configuration $\vec{\sigma}$. In the CDA, we do not have access to $\vec{\sigma}$, but we bypass this by writing an approximate rule that mimics FMS:

\begin{equation}
 r_i^{\text{FMS-CDA}}(\sigma_i, \vec{\sigma}_{\partial i}) = \frac{E_i}{K \langle E\rangle } \, \text{min}\{1, \eta^{\Delta E} \} \label{eq:rate_FMS_CDA}
\end{equation}

We substitute $E$ for its average $\langle E \rangle$, which we can compute using local probabilities $P_t(\vec{\sigma}_a)$. Then, everything is set to run CDA (Eq. \ref{eq:CDA1}) using the FMS dynamic rules.

{\bf G-WalkSAT} is a bit more involved. First, it selects an unsatisfied clause uniformly at random; then, it chooses one of two options: i) with probability $q$ takes a variable inside the clause uniformly at random, and ii) with probability $1-q$ takes the variable belonging to the smaller number of satisfied clauses. The first is known as a random step, and the second is a greedy step. The goal of the latter is not to touch too many satisfied clauses. The algorithmic parameter $q \in [0, 1]$ is also interpretable as the level of noise during dynamics. Note that in both algorithms $q=1$ and $\eta=1$ correspond to the random WalkSAT \cite{PapadimitrouWalkSAT}. 

In this case, to write the exact form of $r_i^{\text{GW}}$, one needs to know the number of satisfied clauses containing $\sigma_i$ and the number of satisfied clauses containing each of its neighbors. This is inconvenient for the implementation of the CDA, and we make a simplification to write:

\begin{equation}
 r_i^{\text{GW-CDA}}(\sigma_i, \sigma_{\partial i}) = (1 - q) \, \frac{E_i}{K\langle E \rangle} + q \frac{E_i}{\langle E \rangle} p(\text{g} | S) \label{eq:rate_GW_CDA}
\end{equation}
where $p(\text{g} | S)$ is the probability of choosing a variable in a clause, given the algorithm decided to take a greedy step and that the variable belongs to $S$ satisfied clauses. If we define $p(S' = S)$ as the probability of finding a neighbor belonging to the same number $S$ of satisfied clauses, and $p(S' > S)$ as the probability that this number is larger than $S$, we can write the following:

\begin{equation}
 p(\text{g} | S) = \sum_{z=0}^{K - 1} \binom{K - 1}{z} \frac{[p(S' = S)]^{z}}{z + 1}  \,  [p(S' > S)]^{K - 1 - z} \label{eq:p_flip_greedy}
\end{equation}

For simplicity, in Eq. \ref{eq:p_flip_greedy} we assume that all the neighbors of the variable have the same probabilities $p(S' = S)$ and $p(S' > S)$. To complete our rule, we compute $p(S'=S)$ averaging over the neighbor's connectivity, assuming that the probability for any other clause to be satisfied is well described by the average $p_{\text {sat}} = 1-\langle E \rangle / M$. The reader can find the details in the appendixes. The result is that $p(S'=S)$ is a Poisson distribution with mean $\alpha K (1 - \langle E \rangle / M)$. Then, $p(S' > S)$ is related to the cumulative of the same Poisson distribution, and we can get them both without much computational effort. With this, we are also ready to run the CDA for the dynamics of G-WalkSAT.

\section{Results}

\begin{figure}[H]
\centering
\includegraphics[width=.8\linewidth]{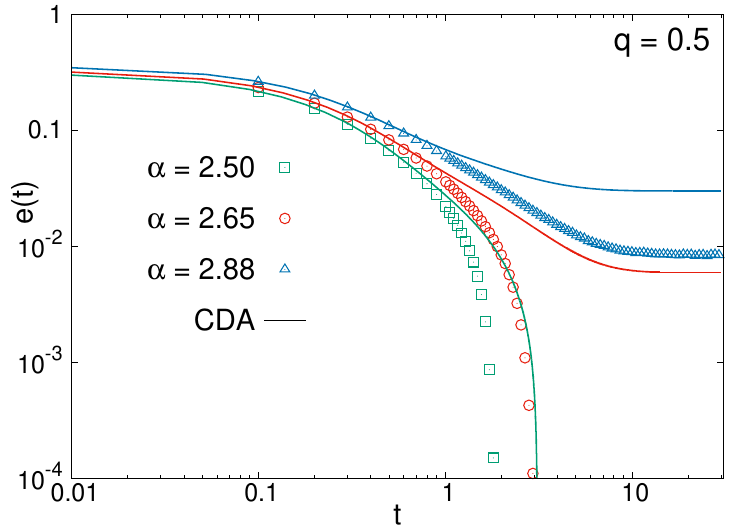}

\includegraphics[width=.8\linewidth]{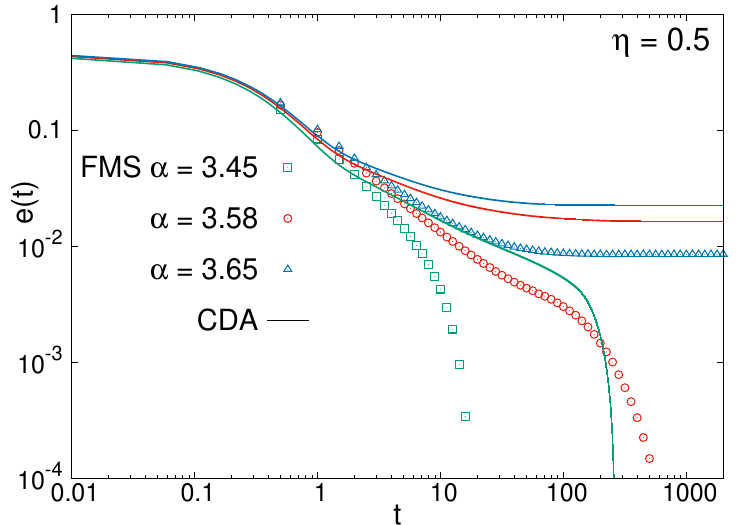}
    \caption{Algorithmic dynamics of G-WalkSAT (top panel) and FMS (bottom panel) in the random 3-SAT. Both panels show the time evolution of the energy density for different values of $\alpha$ in logarithmic scale. The variables are initially assigned to be $0$ or $1$ with the same probability. Points represent an average over $1000$ runs of the algorithm for a single 3-SAT formula. Lines are the prediction of the CDA for the algorithmic dynamics on the same formulas. System size is $N=5 \times 10^{4}$ in all cases. {\bf a)} G-WalkSAT with $q=0.5$. {\bf b)} FMS with $\eta=0.5$.}
    \label{fig:alg_dyn}
\end{figure}

Fig. \ref{fig:alg_dyn} illustrates the temporal evolution of the energy density for different values of $\alpha$, comparing simulations of the FMS and G-WalkSAT algorithms with theoretical predictions from the CDA framework. CDA exhibits a behavior qualitatively similar to both algorithms. For small $\alpha$, the energy density curves display downward curvatures at long times on a logarithmic scale, indicative of polynomial-time convergence to solutions. Conversely, at large $\alpha$, the energy density saturates, reflecting a transition to exponential scaling in solution times. The critical value $\alpha_{\textrm{alg}}$, which demarcates the boundary between polynomial and non-polynomial regimes, defines the algorithmic threshold beyond which the problem becomes computationally intractable by these heuristics.  

%In Fig. \ref{fig:alg_dyn}, we show the evolution in time of the energy density for different values of $\alpha$. We compare the dynamics of FMS and G-WalkSAT with the predictions obtained by means of the numerical integration of CDA. We observe similar behaviors in both cases. For low values of $\alpha$, the density energy curves down for long times. Considering that we are using a logarithmic scale, this type of curvature indicates convergence to solutions in polynomial times. For large $\alpha$ the curves saturate signaling the regime where the time to reach a solution is no longer polynomial. The value $\alpha_{\textrm{alg}}$ that separates these two regimes marks the algorithmic threshold. 

However, notice that the threshold predicted by CDA $\alpha_{\textrm{CDA}}$ is lower than the one predicted by both algorithms. In the upper panel of Fig. \ref{fig:alg_dyn}, the real G-WalkSAT's transition is located between $\alpha=2.65$ and $\alpha=2.88$, while CDA predicts a lower value between $\alpha=2.50$ and $\alpha=2.65$. However, the time scale of the last observed convergence (red points for G-WalkSAT and green line for CDA) coincides very well. At the first $\alpha$ where they both stop converging to zero (blue points for G-WalkSAT and red line for CDA), the energy densities reached at the steady state are also of the same order. The same qualitative picture is obtained for FMS (see the bottom panel of Fig. \ref{fig:alg_dyn}), even though the shape of the curves, the values of $\alpha$, and the time scales to convergence are entirely different if compared with G-WalkSAT. We corroborated that this behavior extends to other values of $q$ and $\eta$. Similar plots, using different parameters, are presented in our appendixes.

\begin{figure}[H]
\centering   \includegraphics[width=.8\linewidth]{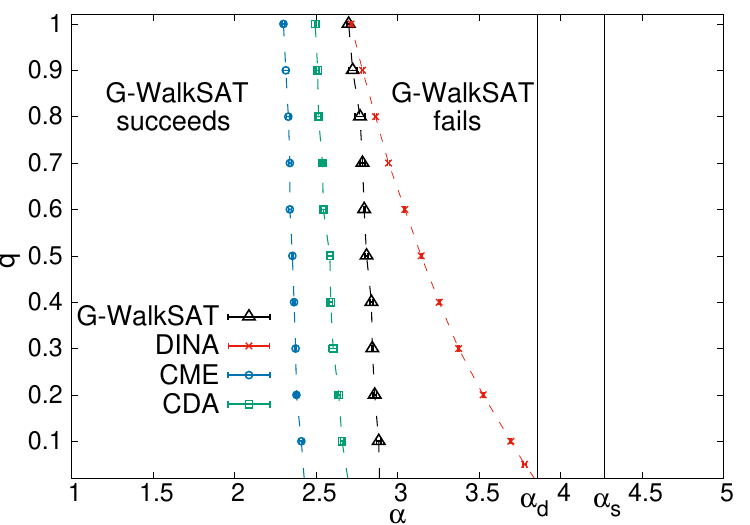}

\includegraphics[width=.8\linewidth]{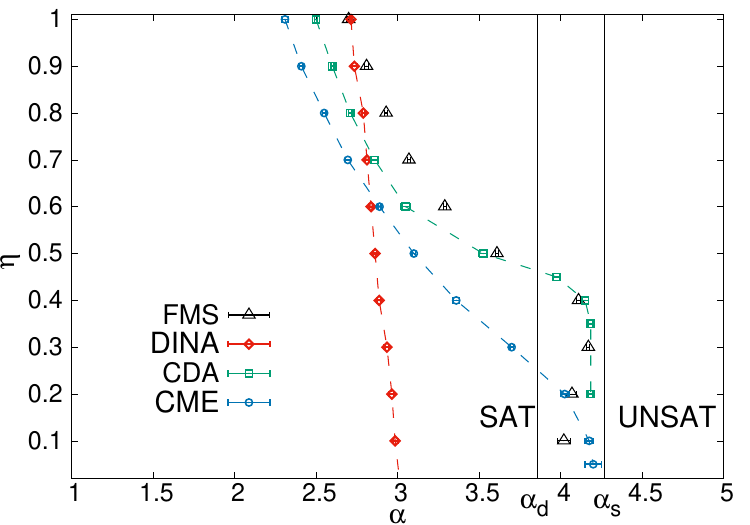}
    \caption{Phase diagrams for the algorithms (triangles) in the random 3-SAT, together with predictions of DINA, CME, and CDA. Both panels show the thresholds for several values of the algorithmic parameter ($q$ for G-WalkSAT and $\eta$ for FMS). To obtain them, we plot energy \textit{vs.} time and observe the curvature at long times. The transition is defined at the $\alpha$ where the type of curvature changes (see Fig. \ref{fig:alg_dyn}). Vertical lines mark the dynamical transition $\alpha_d \approx 3.86$ and the SAT-UNSAT transition $\alpha_s \approx 4.267$. \textbf{a)} G-WalkSAT was run in formulas with $N=5 \times 10^{4}$ variables. For the CDA and the CME, the system size was $N=10^{5}$ for $q \geq 0.2$ and $N=5 \times 10^{4}$ for $q\leq0.1$. \textbf{b)} FMS was run in formulas with $N=5 \times 10^{4}$ variables for $\eta \geq 0.6$, and with $N=\times 10^{5}$ variables for $\eta \leq 0.5$. For the CDA and the CME, we used $N=5 \times 10^{4}$, except for the last two points of the CME that were obtained with $N=5 \times 10^{5}$.}
    \label{fig:phase_diagrams}
\end{figure}

Observing the curvatures at different values of $\alpha$, we can determine the location of the threshold $\alpha_{\textrm{alg}}$ with sufficient precision. As was pointed out before \cite{Seitz_2005}, this threshold varies as a function of the corresponding algorithmic parameter. The dependence of $\alpha_{\textrm{alg}}$ on $q$ or $\eta$ is presented in the phase diagrams of Fig. \ref{fig:phase_diagrams}. The continuous vertical lines represent the theoretical predictions derived from the cavity method \cite{MertensKSAT2006, Montanari_2008}, while the algorithmic thresholds of G-WalkSAT and FMS are marked with triangles. To the left of the triangles, the algorithm finds solutions in polynomial time; to the right, it fails to do so.  

Note that CDA (green squares) describes the results of the algorithms much better than previous approximations in the literature (\textit{i.e.} DINA \cite{BarthelwalkSAT2003} and CME \cite{CME-PRL}). In Ref. \cite{CME-PRL}, CME was applied only to FMS, with some numerical difficulties in predicting the algorithmic threshold for low values of $\eta$. Here, exploiting a fast computational implementation of CDA and CME, we present more accurate results up to $\eta=0.01$. The efficient algorithm for the computation of the derivatives in Eq. \ref{eq:CDA1} is explained in the appendixes.

In Ref. \cite{BarthelwalkSAT2003}, DINA was applied only to G-WalkSAT, and the phase diagram was not reported. For comparison, we include the predicted phase diagrams for G-WalkSAT (in the upper panel of Fig. \ref{fig:phase_diagrams}) and for FMS (in the lower panel). The name we chose, DINA, is taken from the follow-up work in Ref. \cite{Semerjiandyn2004}, where this technique is explained in more detail. For completeness, we include our brief recap of DINA's equations and approximations in the appendixes.

%The bottom panel of Fig. \ref{fig:phase_diagrams} shows that, for $\eta<0.5$, FMS solves instances in polynomial time also inside the dynamical phase estimated from the cavity method (beyond $\alpha_d=3.86$). CDA reproduces this result correctly. Moreover, when $\eta$ is below $0.4$, the CDA predicts a fast convergence to solutions for all $\alpha \leq 4.17$, in good agreement with the true algorithmic threshold.  

For all values of $q$ of the G-WalkSAT algorithm and for $\eta > 0.5$ in FMS, the already mentioned \textit{shift} between CDA's predictions and the observed algorithmic thresholds is independent of $q$ or $\eta$. This means that away from the influence of the dynamical transition at $\alpha_d$, the distance $\alpha_{\textrm{alg}}-\alpha_{\textrm{CDA}}$ does not strongly depend on the algorithmic rules. In this zone, we expect that the accuracy of the approximate master equations can be improved by including correlations between variables at longer distances in the graph. The results in Ref. \cite{CDA-Pspin} show that these correlations can be crucial in describing some dynamical behaviors. Moreover, the bottom panel of Fig. \ref{fig:phase_diagrams} shows that, for $\eta<0.5$, FMS solves instances in polynomial time also inside the dynamical phase estimated from the cavity method (beyond $\alpha_d=3.86$). CDA reproduces this result correctly. When $\eta$ is below $0.4$, the CDA predicts a fast convergence to solutions for all $\alpha \leq 4.17$, in good agreement with the true algorithmic threshold.  

Now, by definition, to the left of the green squares in Fig. \ref{fig:phase_diagrams}, the numerical integration of CDA's equations leads to small energies and, as a by-product, one has access to the marginals $P^{\text{CDA}}(\sigma_i)$ for all $i=1,\ldots, N$. A natural question to explore is whether these marginals contain useful information about the local structure of the solutions.

%The natural question is: Are these marginals well-informed about the local structure of the solutions?

%When FMS's parameter $\eta$ is below $0.4$, the CDA predicts a fast convergence to solutions for all $\alpha < 4.17$, in good agreement with the true algorithmic threshold. The numerical integration of these equations leads to small energies and, as a by-product, one has access to the marginals $P^{\text{CDA}}(\sigma_i)$ for all $i=1,\ldots, N$. The natural question is: are these marginals also well-informed about the local structure of the solutions?

To answer this question, we devised a decimation procedure that translates the prediction of the CDA into candidate solutions. While FMS implements a stochastic local search to explore these configurations, the new CDA-guided decimation is a deterministic algorithm with a well-defined running time. We present its pseudo-code here as Algorithm \ref{alg:CDA-d}.

The idea is to make $N$ consecutive reductions of the original formula, fixing a single variable each time, until we obtain a final candidate solution $\vec{\sigma}^{\ast}$. The parameter $\tau$ is the number of integration steps of the CDA performed between two consecutive reductions. The transition rules $r_i(\sigma_i, \vec{\sigma}_{\partial i})$ are also received as input parameters and can be changed to mimic different algorithmic dynamics. In this work, we use FMS's rules (see Eq. \ref{eq:rate_FMS_CDA}).

\begin{algorithm}[H]
\begin{algorithmic}[1]
    \State{{\bf input} Boolean formula with $N$ variables and $M$ clauses}
    \State{Choose a positive integer $\tau$}
    \State{Choose the dynamic rules $r_i(\sigma_i, \vec{\sigma}_{\partial i})$}
    \State{Set random initial probabilities for CDA}
    \For{$t=1, \ldots, N$}
    \State{Take $\tau$ steps of the numerical integrator of the CDA}
    \State{Compute the magnetizations $m_i^{\text{CDA}}$}
    \State{Find $i$ with the largest $|m_i^{\text{CDA}}|$}
    \State{{\bf if} $m_i^{\text{CDA}} > 0$ {\bf then} set $\sigma_i^{\ast}=1$}
    \State{{\bf else} set $\sigma_i^{\ast}=-1$}
    \State{Make the corresponding reduction to the formula.}
    \EndFor \\
    \Return{Final assignment $\vec{\sigma}^{\ast}$}
  \end{algorithmic}
 \caption{CDA-guided decimation}
 \label{alg:CDA-d}
\end{algorithm}

After $\tau$ and $r_i$ are set, the key is to decide which variable to fix each time and to what value. For this purpose, we used the local magnetizations $m_i^{\text{CDA}} \equiv \langle \sigma_i \rangle = \sum_{\sigma_i} \sigma_i \, P_i^{\text{CDA}}(\sigma_i)$ predicted by the CDA. When we get a magnetization $m_i^{\text{CDA}}$ close to $1$, we assume this as an indication that in most FMS's solutions the $i$-th variable takes the value $\sigma_i=1$. Similarly, $m_i^{\text{CDA}}$ close to $-1$ would indicate that $\sigma_i=-1$ in most cases. At each reduction, we choose the magnetization $m_i^{\text{CDA}}$ with the highest absolute value and fix the corresponding variable in the direction suggested by $m_i^{\text{CDA}}$.

\begin{figure}%[tbhp]
\centering
\includegraphics[width=.8\linewidth]{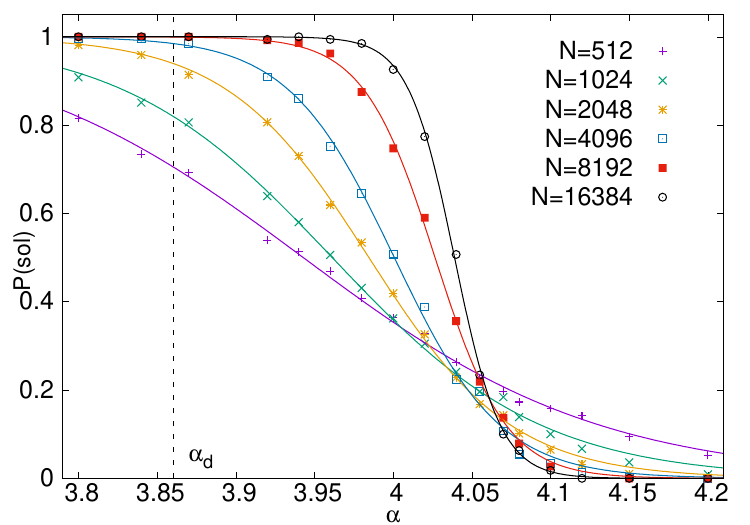}
    \caption{CDA-guided decimation with FMS rates in 3-SAT formulas for various system sizes $N$. Points represent the fraction of the instances solved for different values of $\alpha$. The variables were decimated one at a time, taking $\tau=5$ steps of the numeric integrator between consecutive reductions of the formula. The statistics include $1000$ formulas for each pair $(\alpha, N)$. The lines are logistic fits to the points.}
    \label{fig:CDA-guided_decimation}
\end{figure}

The results of this decimation are presented in Fig. \ref{fig:CDA-guided_decimation}. We show that the CDA-guided decimation with FMS's transition rules is capable of finding solutions well beyond the dynamical transition $\alpha_d$, revealing that CDA's marginals $P^{\text{CDA}}(\sigma_i)$ are indeed informative about the local structure of solutions inside the supposedly \emph{hard} region ($\alpha_d < \alpha<\alpha_s$). Fig. \ref{fig:CDA-guided_decimation} contains results obtained with $\tau=5$ for several system sizes. The curves become steeper as the system size $N$ increases. To the left of $\alpha \sim 4.05$, the probability of finding a solution increases with $N$; to the right, the probability decreases for larger $N$. This indicates an algorithmic transition around $\alpha\sim4.05$, definitely larger than $\alpha_d=3.86$.

The reader should notice two important advantages of CDA-guided decimation. First, it notably outperforms the well-known Belief Propagation (BP) with decimation, which remains blocked at $\alpha < \alpha_d$ in the random 3-SAT \cite{mezard2009information}. While BP is a technique designed to compute the equilibrium distribution, hence operating with the Gibbs-Boltzmann measure, the CDA can be applied to non-equilibrium measures. With this comes the second important advantage: in Fig. \ref{fig:CDA-guided_decimation} we used the FMS's transition rules for the CDA's integration, but nothing forbids new applications from trying other dynamic rules that could improve the performance of the decimation procedure. We can also vary another hyperparameter in CDA-guided decimation: the number $\tau$ of steps between consecutive reductions of the formula. In the appendixes, we show that by increasing $\tau$ we achieve better algorithmic performances.\footnote{The code necessary to reproduce all the figures in this section is available at {\color{blue} \url{https://github.com/d4v1d-cub/ApproxMasterEqKSAT.git}}} 

\section{Discussion}

Theoretical approaches for characterizing the energy landscape of hard combinatorial optimization problems, like random K-SAT, are typically intricate, reflecting the inherent complexity of these landscapes. For small $K$, the most accurate method available to approach this problem is the cavity method \cite{mezardksat2002}, which predicts that above the clustering transition $\alpha_d$ the uniform measure over solutions develops long-range correlations and fractures into exponentially many clusters \cite{KrzakalaKSAT2007}. 

In contrast, our Conditional Dynamics Approximation (CDA) relies solely on local correlations to model algorithmic dynamics, yet this simplification suffices to show that the algorithmic threshold of G-WalkSAT is almost independent of $q$ and to predict the behavior of Focused Metropolis Search (FMS) beyond $\alpha_d$, including its algorithmic threshold. Furthermore, we demonstrate that CDA’s marginals can be transformed into concrete solutions via decimation, revealing that they encode meaningful information about the local structure of FMS’s solution space and opening the path to explore the potential of other non-equilibrium measures.

%The techniques applied to describe the energy landscape in hard combinatorial optimization problems such as random K-SAT are usually involved, which is necessary to match the complexity of the landscapes. When $K$ is small, the best available approach is the cavity method \cite{mezardksat2002}. It predicts that, above $\alpha_d$, the uniform measure of solutions displays long-range correlations and breaks into an exponential number of clusters \cite{KrzakalaKSAT2007}. In contrast to this picture, CDA includes only local correlations to describe specific algorithmic dynamics, and this ingredient is enough to make predictions about the FMS dynamics beyond $\alpha_d$, including an algorithmic threshold for FMS close to the true optimal threshold. Moreover, we show that CDA's marginals van be translated into actual solutions using decimation, indicating that it gets information about the local structure of FMS's solutions.

Our results are not in contradiction with the predictions of the cavity method. The key insight is that FMS and G-WalkSAT operate far from equilibrium and do not sample solutions according to the uniform measure. Previous works \cite{BaldassiJSTAT2016, BaldassiPNAS2016} suggest that exploring dense solution clusters within highly entropic regions can result in algorithmic thresholds surpassing $\alpha_d$. In contrast, the accurate description of FMS beyond $\alpha_d$ by CDA, which uses only local correlations, suggests that the solutions in the solution space identified in \cite{BaldassiJSTAT2016, BaldassiPNAS2016} can then be well characterized using local statistics alone. A unique advantage of CDA is its direct access to the local marginals of the stationary distribution, which provides a powerful tool to probe these solution-dense regions. Current work is leveraging this capability.

, %without paying the price of replicating the systems as in Refs. \cite{BaldassiPNAS2016, MariaChiara2023PRX}. We plan to explore these and other possible directions of work in the future.

\section*{Acknowledgments}
We thank E Aurell for early discussions on this problem and Maria Chiara Angelini, Alfredo Braunstein, and Daniel Estévez for fruitful insights and suggestions. This work has been funded by European Union - NextGenerationEU by the project PRIN 2022 PNRR, P20229PBZR. This study was conducted using the DARIAH HPC cluster at CNR-NANOTEC in Lecce, funded by the "MUR PON Ricerca e Innovazione 2014-2020" project, code PIR01\_00022.

\bibliographystyle{unsrt}
\bibliography{ref_cda_2025} % Replace "references" with the name of your bibliography file

\begin{thebibliography}{10}

\bibitem{mezardksat2002}
M.~Mezard, G.~Parisi, and R.~Zecchina.
\newblock Analytic and algorithmic solution of random satisfiability problems.
\newblock {\em Science}, 297(5582):812--815, 2002.

\bibitem{MertensKSAT2006}
Stephan Mertens, Marc M\'ezard, and Riccardo Zecchina.
\newblock Threshold values of random k-sat from the cavity method.
\newblock {\em Random Structures \& Algorithms}, 28(3):340--373, 2006.

\bibitem{KrzakalaKSAT2007}
Florent Krzakala, Andrea Montanari, Federico Ricci-Tersenghi, Guilhem
  Semerjian, and Lenka Zdeborov\'a.
\newblock Gibbs states and the set of solutions of random constraint
  satisfaction problems.
\newblock {\em Proceedings of the National Academy of Sciences},
  104(25):10318--10323, 2007.

\bibitem{Montanari_2008}
Andrea Montanari, Federico Ricci-Tersenghi, and Guilhem Semerjian.
\newblock Clusters of solutions and replica symmetry breaking in random
  k-satisfiability.
\newblock {\em Journal of Statistical Mechanics: Theory and Experiment},
  2008(04):P04004, apr 2008.

\bibitem{SemerjianwalkSAT2003}
Guilhem Semerjian and R\'emi Monasson.
\newblock Relaxation and metastability in a local search procedure for the
  random satisfiability problem.
\newblock {\em Physical Review E}, 67:066103, 2003.

\bibitem{Semerjiandyn2004}
G.~Semerjian and M.~Weigt.
\newblock Approximation schemes for the dynamics of diluted spin models: the
  ising ferromagnet on a bethe lattice.
\newblock {\em Journal of Physics A: Mathematical and General},
  37(21):5525--5546, may 2004.

\bibitem{ArdeliusASAT2006}
John Ardelius and Erik Aurell.
\newblock Behavior of heuristics on large and hard satisfiability problems.
\newblock {\em Physical Review E}, 74:037702, Sep 2006.

\bibitem{AlavaPNAS2008}
Mikko Alava, John Ardelius, Erik Aurell, Petteri Kaski, Supriya Krishnamurthy,
  Pekka Orponen, and Sakari Seitz.
\newblock Circumspect descent prevails in solving random constraint
  satisfaction problems.
\newblock {\em Proceedings of the National Academy of Sciences},
  105(40):15253--15257, 2008.

\bibitem{Lage2014message}
A~Lage-Castellanos, R~Mulet, and F~Ricci-Tersenghi.
\newblock Message passing and monte carlo algorithms: Connecting fixed points
  with metastable states.
\newblock {\em EPL (Europhysics Letters)}, 107(5):57011, 2014.

\bibitem{LemoyVFMS2015}
R\'emi Lemoy, Mikko Alava, and Erik Aurell.
\newblock Local search methods based on variable focusing for random
  $k$-satisfiability.
\newblock {\em Physical Review E}, 91:013305, Jan 2015.

\bibitem{Monasson2004}
Remi Monasson, Simona Cocco, Guilhem Semerjian, and Andrea Montanari.
\newblock Approximate analysis of search algorithms with ``physical'' methods.
\newblock In {\em Phase Transitions and Algorithmic Complexity}, pages 1--20.
  Santa Fe Institute, 2004.

\bibitem{Achlioptas08}
Dimitris Achlioptas and Amin Coja-Oghlan.
\newblock Algorithmic barriers from phase transitions.
\newblock In {\em 2008 49th Annual IEEE Symposium on Foundations of Computer
  Science}, pages 793--802, 2008.

\bibitem{Budzynski_2019}
Louise Budzynski, Federico Ricci-Tersenghi, and Guilhem Semerjian.
\newblock Biased landscapes for random constraint satisfaction problems.
\newblock {\em Journal of Statistical Mechanics: Theory and Experiment},
  2019(2):023302, feb 2019.

\bibitem{AurellNIPS2004WalkSAT}
Erik Aurell, Uri Gordon, and Scott Kirkpatrick.
\newblock Comparing beliefs, surveys, and random walks.
\newblock In L.~Saul, Y.~Weiss, and L.~Bottou, editors, {\em Advances in Neural
  Information Processing Systems}, volume~17. MIT Press, 2004.

\bibitem{Seitz_2005}
Sakari Seitz, Mikko Alava, and Pekka Orponen.
\newblock Focused local search for random 3-satisfiability.
\newblock {\em Journal of Statistical Mechanics: Theory and Experiment},
  2005(06):P06006, jun 2005.

\bibitem{achlioptas2006solution}
Dimitris Achlioptas and Federico Ricci-Tersenghi.
\newblock On the solution-space geometry of random constraint satisfaction
  problems.
\newblock In {\em Proceedings of the thirty-eighth annual ACM symposium on
  Theory of computing}, pages 130--139, 2006.

\bibitem{ZdeborovaPRE2007qcol}
Lenka Zdeborov\'a and Florent Krzakala.
\newblock Phase transitions in the coloring of random graphs.
\newblock {\em Phys. Rev. E}, 76:031131, Sep 2007.

\bibitem{gamarnik2017limits}
David Gamarnik and Madhu Sudan.
\newblock Limits of local algorithms over sparse random graphs.
\newblock {\em Annals of probability: An official journal of the Institute of
  Mathematical Statistics}, 45(4):2353--2376, 2017.

\bibitem{gamarnik2018finding}
David Gamarnik and Quan Li.
\newblock Finding a large submatrix of a gaussian random matrix.
\newblock {\em The Annals of Statistics}, 46(6A):2511--2561, 2018.

\bibitem{gamarnik2021overlap}
David Gamarnik.
\newblock The overlap gap property: A topological barrier to optimizing over
  random structures.
\newblock {\em Proceedings of the National Academy of Sciences},
  118(41):e2108492118, 2021.

\bibitem{BaldassiJSTAT2016}
Carlo Baldassi, Alessandro Ingrosso, Carlo Lucibello, Luca Saglietti, and
  Riccardo Zecchina.
\newblock Local entropy as a measure for sampling solutions in constraint
  satisfaction problems.
\newblock {\em Journal of Statistical Mechanics: Theory and Experiment},
  2016(2):023301, feb 2016.

\bibitem{BaldassiPNAS2016}
Carlo Baldassi, Christian Borgs, Jennifer~T. Chayes, Alessandro Ingrosso, Carlo
  Lucibello, Luca Saglietti, and Riccardo Zecchina.
\newblock Unreasonable effectiveness of learning neural networks: From
  accessible states and robust ensembles to basic algorithmic schemes.
\newblock {\em Proceedings of the National Academy of Sciences},
  113(48):E7655--E7662, 2016.

\bibitem{MariaChiara2023PRX}
Maria~Chiara Angelini and Federicco Ricci-Tersenghi.
\newblock Limits and performances of algorithms based on simulated annealing in
  solving sparse hard inference problems.
\newblock {\em Physical Review X}, 13:021011, 2023.

\bibitem{BarthelwalkSAT2003}
Wolfgang Barthel, Alexander~K. Hartmann, and Martin Weigt.
\newblock Solving satisfiability problems by fluctuations: The dynamics of
  stochastic local search algorithms.
\newblock {\em Physical Review E}, 67:066104, 2003.

\bibitem{CDA-Pspin}
David Machado, Roberto Mulet, and Federico Ricci-Tersenghi.
\newblock Improved mean-field dynamical equations are able to detect the
  two-step relaxation in glassy dynamics at low temperatures.
\newblock {\em Journal of Statistical Mechanics: Theory and Experiment},
  2023(12):123301, dec 2023.

\bibitem{CME-Pspin}
E.~Aurell, E.~Dom\'{\i}nguez, D.~Machado, and R.~Mulet.
\newblock Exploring the diluted ferromagnetic p-spin model with a cavity master
  equation.
\newblock {\em Physical Review E}, 97:05103(R), 2018.

\bibitem{CME-PRL}
E.~Aurell, E.~Dom\'{\i}nguez, D.~Machado, and R.~Mulet.
\newblock A theory of non-equilibrium local search on random satisfaction
  problems.
\newblock {\em Physical Review Letters}, 123:230602, 2019.

\bibitem{hCME}
D.~Machado and R.~Mulet.
\newblock From random point processes to hierarchical cavity master equations
  for stochastic dynamics of disordered systems in random graphs: Ising models
  and epidemics.
\newblock {\em Physical Review E}, 104:054303, Nov 2021.

\bibitem{vanKampen92}
NG~van Kampen.
\newblock {\em Stochastic Processes in Physics and Chemistry}, volume~1.
\newblock Elsevier, Amsterdam, 1992.

\bibitem{cator2012second}
Eric Cator and Piet Van~Mieghem.
\newblock Second-order mean-field susceptible-infected-susceptible epidemic
  threshold.
\newblock {\em Physical review E}, 85(5):056111, 2012.

\bibitem{mata2013pair}
Ang{\'e}lica~S Mata and Silvio~C Ferreira.
\newblock Pair quenched mean-field theory for the
  susceptible-infected-susceptible model on complex networks.
\newblock {\em EPL (Europhysics Letters)}, 103(4):48003, 2013.

\bibitem{pastor2015epidemic}
Romualdo Pastor-Satorras, Claudio Castellano, Piet Van~Mieghem, and Alessandro
  Vespignani.
\newblock Epidemic processes in complex networks.
\newblock {\em Reviews of Modern Physics}, 87(3):925, 2015.

\bibitem{silva19PQMF}
Diogo~H. Silva, Silvio~C. Ferreira, Wesley Cota, Romualdo Pastor-Satorras, and
  Claudio Castellano.
\newblock Spectral properties and the accuracy of mean-field approaches for
  epidemics on correlated power-law networks.
\newblock {\em Physical Review Research}, 1:033024, Oct 2019.

\bibitem{silva20PQMF}
Diogo~H. Silva, Francisco~A. Rodrigues, and Silvio~C. Ferreira.
\newblock High prevalence regimes in the pair-quenched mean-field theory for
  the susceptible-infected-susceptible model on networks.
\newblock {\em Physical Review E}, 102:012313, Jul 2020.

\bibitem{CME-AvAndBP}
E.~Dom\'{\i}nguez, D.~Machado, and R.~Mulet.
\newblock The cavity master equation: average and fixed point of the
  ferromagnetic model in random graphs.
\newblock {\em Journal of Statistical Mechanics: Theory and Experiment},
  2020:073304, 2020.

\bibitem{Metropolis1953}
Nicholas Metropolis, Arianna~W. Rosenbluth, Marshall~N. Rosenbluth, Augusta~H.
  Teller, and Edward Teller.
\newblock Equation of state calculations by fast computing machines.
\newblock {\em The Journal of Chemical Physics}, 21(6):1087--1092, 06 1953.

\bibitem{PapadimitrouWalkSAT}
CH~Papadimitriou.
\newblock On selecting a satisfying truth assignment.
\newblock In New~York IEEE Computer~Society, editor, {\em Proceedings of the
  32nd IEEE Symposium on the Foundations of Computer Science}, volume FOCS-91,
  page 163–169, 1991.

\bibitem{mezard2009information}
Marc Mezard and Andrea Montanari.
\newblock {\em Information, physics, and computation}.
\newblock Oxford University Press, 2009.

\end{thebibliography}

% \end{multicols}

\pagebreak

\onecolumn

\appendix
\section{Average CDA} \label{sec:avCDA1}
The CDA is written for the set of probabilities $P_t(\vec{\sigma}_a)$. In a given random K-SAT formula, there is one probability like these for each clause. When running in a single graph, each variable $\sigma_i$, with $i \in a$, belongs to a given number $\gamma$ of other clauses. A group of those clauses will be satisfied when $\sigma_i=1$, and some other clauses will be satisfied when $\sigma_i=-1$. Let us define $l^{+}$ and $l^{-}$ as the number of clauses in the first and second groups, respectively. 

To obtain the average results (av-CDA) in the main section, we assume that the prediction of the original CDA is well described by a population of probabilities $P_t(\vec{s} ; \vec{l}^{+}, \vec{l}^{-})$. Each element of the population is a joint probability defined over a clause and their argument is a vector $\vec{s}=\{s_1, \ldots, s_K\}$, with $s_z=0,1$. The component $s_{z}$ ($z=1,\ldots, K$) is '$0$' when the corresponding variable satisfies the clause and is '$1$' otherwise. These probabilities have two associated vectors of parameters $\vec{l}^{+} = \{l_1^{+}, \ldots, l_K^{+} \}$ and $\vec{l}^{-} = \{l_1^{-}, \ldots, l_K^{-} \}$. The integer $l_{z}^{+}$($l_{z}^{-}$) is the number of other clauses that are satisfied when $s_z=0$($s_z=1$).

We can write average equations for those probabilities, starting from the single instance version of the CDA. The readers will find them intuitive if they remember the definition of $P(\vec{s} ; \vec{l}^{+}, \vec{l}^{-})$. The equations are:

\begin{eqnarray}
\frac{dP_t(\vec{s} ; \vec{l}^{+}, \vec{l}^{-})}{dt} = -\sum_{z=1}^{K}(1-2 s_z) \sum_{u^{+}=0}^{l_{z}^{+}} \binom{l_{z}^{+}}{u^{+}} \sum_{u^{-}=0}^{l_{z}^{-}} \binom{l_{z}^{-}}{u^{-}} r \Big(u^{-} , u^{+}+ \prod_{j \neq z}^{K} s_j\Big) P(\vec{s}_{\setminus z}, s_z=0 ; \vec{l}^{+}, \vec{l}^{-}) \times \nonumber \\
\times \Big[ P_{US}(l_z^{+}, l_z^{-}) \Big]^{u^{+}} \Big[ P_{SS}(l_z^{+}, l_z^{-}) \Big]^{l_z^{+}-u^{+}} \Big[ P_{UU}(l_z^{-}-1, l_z^{+}+1) \Big]^{u^{-}} \Big[ P_{SU}(l_z^{-}-1, l_z^{+}+1)\Big]^{l_z^{-}-u^{-}} + \nonumber \\
+ \sum_{z=1}^{K} (1-2 s_z)\sum_{u^{+}=0}^{l_{z}^{+}} \binom{l_{z}^{+}}{u^{+}} \sum_{u^{-}=0}^{l_{z}^{-}} \binom{l_{z}^{-}}{u^{-}} r \big(u^{+}+ \prod_{j\neq z}^{K} s_j , u^{-}\big) P(\vec{s}_{\setminus z}, s_z=1 ; \vec{l}^{+}, \vec{l}^{-}) \times \nonumber \\
\times \Big[ P_{UU}(l_z^{+}, l_z^{-}) \Big]^{u^{+}} \Big[ P_{SU}(l_z^{+}, l_z^{-}) \Big]^{l_z^{+}-u^{+}} \Big[ P_{US}(l_z^{-}-1, l_z^{+}+1) \Big]^{u^{-}} \Big[ P_{SS}(l_z^{-}-1, l_z^{+}+1)\Big]^{l_z^{-}-u^{-}}
 \label{eq:avCDA1}
\end{eqnarray}
where $s_z$, $l_z^{+}$, and $l_z^{-}$ are the components of the vectors $\vec{s}$, $\vec{l}^{+}$, and $\vec{l}^{-}$, respectively. The vector $\vec{s}_{\setminus z}$ contains all the variables $s_j$, with $j=1,\ldots, K$, such that $j \neq z$. The parameter $u^{-}$ is the number of unsatisfied clauses in the neighborhood of the variable when $s_z=0$. On the other hand, $u^{+} + \prod_{j\neq z}^{K} s_j $ is the number of unsatisfied clauses when $s_z=1$. Notice that the product $\prod_{j\neq z}^{K} s_j$ is one only if all the variables inside $\vec{s}_{\setminus z}$ are not satisfying the clause. 

To write Eq. \ref{eq:avCDA1}, we consider transition rates $r$ that depend, at most, on the current number of unsatisfied clauses containing the variable and the number of unsatisfied clauses after flipping it. Both, the rates of G-WalkSAT and FMS, fit into this category. The expression for $r_i$ can be seen in Eqs. \ref{eq:rate_FMS_av} and \ref{eq:rate_GW_av}. With this, we just need to define the four conditional probabilities in the second and fourth lines of Eq. \ref{eq:avCDA1}. These are:

\begin{eqnarray}
    P_{UU}(l^{+}, l^{-}) = \Big[ \sum_{\vec{l}^{+}}\sum_{\vec{l}^{-}} \sum_{\vec{s}} P(\vec{s}; \vec{l}^{+}, \vec{l}^{-}) \delta(l_1^{+}, l^{+}) \delta(l_1^{-}, l^{-}) \delta(s_1,1) \Big]^{-1} \sum_{\vec{l}^{+}}\sum_{\vec{l}^{-}} \sum_{\vec{s}} P(\vec{s}; \vec{l}^{+}, \vec{l}^{-}) \delta(l_1^{+}, l^{+}) \delta(l_1^{-}, l^{-}) \delta\Big(\prod_{j=2}^{K} s_j, 1\Big) \delta(s_1,1) \nonumber\\
    P_{SU}(l^{+}, l^{-}) = \Big[ \sum_{\vec{l}^{+}}\sum_{\vec{l}^{-}} \sum_{\vec{s}} P(\vec{s}; \vec{l}^{+}, \vec{l}^{-}) \delta(l_1^{+}, l^{+}) \delta(l_1^{-}, l^{-}) \delta(s_1,1) \Big]^{-1} \sum_{\vec{l}^{+}}\sum_{\vec{l}^{-}} \sum_{\vec{s}} P(\vec{s}; \vec{l}^{+}, \vec{l}^{-}) \delta(l_1^{+}, l^{+}) \delta(l_1^{-}, l^{-}) \delta\Big(\prod_{j=2}^{K} s_j, 0\Big) \delta(s_1,1) \nonumber\\
    P_{US}(l^{+}, l^{-}) = \Big[ \sum_{\vec{l}^{+}}\sum_{\vec{l}^{-}} \sum_{\vec{s}} P(\vec{s}; \vec{l}^{+}, \vec{l}^{-}) \delta(l_1^{+}, l^{+}) \delta(l_1^{-}, l^{-}) \delta(s_1,0) \Big]^{-1} \sum_{\vec{l}^{+}}\sum_{\vec{l}^{-}} \sum_{\vec{s}} P(\vec{s}; \vec{l}^{+}, \vec{l}^{-}) \delta(l_1^{+}, l^{+}) \delta(l_1^{-}, l^{-}) \delta\Big(\prod_{j=2}^{K} s_j, 1\Big) \delta(s_1,0) \nonumber\\
    P_{SS}(l^{+}, l^{-}) = \Big[ \sum_{\vec{l}^{+}}\sum_{\vec{l}^{-}} \sum_{\vec{s}} P(\vec{s}; \vec{l}^{+}, \vec{l}^{-}) \delta(l_1^{+}, l^{+}) \delta(l_1^{-}, l^{-}) \delta(s_1,0) \Big]^{-1} \sum_{\vec{l}^{+}}\sum_{\vec{l}^{-}} \sum_{\vec{s}} P(\vec{s}; \vec{l}^{+}, \vec{l}^{-}) \delta(l_1^{+}, l^{+}) \delta(l_1^{-}, l^{-}) \delta\Big(\prod_{j=2}^{K} s_j, 0\Big) \delta(s_1,0)\nonumber
\end{eqnarray}
where $\delta(a,b)$ is a Kronecker delta.

All these conditional probabilities are defined for the configurations of the variables inside a clause. They are always conditioned on the value of one of the variables. Their interpretation is clear from the previous equations:
\begin{itemize}
    \item $P_{UU}(l^{+}, l^{-})$ is the probability that the clause is unsatisfied, given that one of the variables is already not satisfying the clause. That variable $s$ satisfies $l^{+}$ other clauses when $s=0$, and other $l^{-}$ clauses when $s=1$.
    \item $P_{SU}(l^{+}, l^{-})$ is the probability that the clause is satisfied, given that one of the variables is already not satisfying the clause. That variable $s$ satisfies $l^{+}$ other clauses when $s=0$, and other $l^{-}$ clauses when $s=1$.
    \item $P_{US}(l^{+}, l^{-})$ is the probability that the rest of the variables do not satisfy the clause, given that one variable is already satisfying the clause. That variable $s$ satisfies $l^{+}$ other clauses when $s=0$, and other $l^{-}$ clauses when $s=1$.
    \item $P_{SS}(l^{+}, l^{-})$ is the probability that the rest of the variables satisfy the clause, given that one variable already satisfies the clause. That variable $s$ satisfies $l^{+}$ other clauses when $s=0$, and other $l^{-}$ clauses when $s=1$.
\end{itemize}

These equations are in principle solvable if one has all the probabilities $P(\vec{s}; \vec{l}^{+}, \vec{l}^{-})$, but in practice this is impossible. Instead, we run the equations over a population of probabilities with a finite number of elements. We need to make sure, however, that for every pair $(l^{+}, l^{-})$, the pair $(l^{-}-1, l^{+}+1)$ is also present in the population. This is a consequence of the fact that both appear simultaneously in the second and fourth lines of Eq. \ref{eq:avCDA1}.

We thus need a final trick, which is the following. To introduce a probability into the population, we draw each $(l_{z}^{+}, l_{z}^{-})$, with $z=1,\ldots, K$, from the right Poisson distribution. Afterward, we insert also an element with the pairs $(l_z^{-}-1, l_z^{+}+1)$, but with a reweight $w(\vec{l}^{+})$ so that averages over the population have the right form.

In order to explain this clearly, let us take a hypothetical population whose elements $\{(x_i;l^{+}, l^{-})\}$, with $i=1, \ldots, 2m$, are vectors of three elements. Two are inside the associated pair $(l^{+}, l^{-})$ and one is a real number $x_i$. The extension to our case, where $P_i(\vec{s}; \vec{l}^{+}, \vec{l}^{-})$ depends on $K$ distinct pairs $(l_z^{+}, l_z^{-})$ and on a vector $\vec{s}$, will be straightforward.

Each element with an odd index $x_{2k-1}$ is inserted after extracting the number $\gamma_{2k-1}$ from the Poisson distribution: 

\begin{equation}
    Q(\gamma) = e^{-\lambda} \frac{\lambda^{\gamma}}{(\gamma)!}
\end{equation}

The value of $l_{2k-1}^{+}$ is drawn from the binomial;

\begin{equation}
    B(l^{+}\mid \gamma) = \binom{\gamma}{l^{+}} \Big(\frac{1}{2}\Big)^{\gamma}
\end{equation}
and $l_{2k-1}^{-}$ is set to $l_{2k-1}^{-}=\gamma_{2k-1}-l_{2k-1}^{+}$. Then, the value of $x_{2k-1}$ is set to some number $x_0$, independent of $k$.

To insert the element in the position $2k$, we take $l_{2k}^{+}=l_{2k-1}^{-}-1$ and $l_{2k}^{-}=l_{2k-1}^{+}+1$, if $l_{2k-1}^{-} \geq 1$. When $l_{2k-1}^{-}=0$, we set $l_{2k}^{+}=l_{2k-1}^{+}$ and $l_{2k}^{-}=l_{2k-1}^{-}$. We then assign to $x_{2k}$ the value $x_{2k}=w(l_{2k-1}^{+} \mid \gamma_{2k-1}) x_0$. The form of thee reweighting $w(l_{2k-1}^{+} \mid \gamma_{2k-1} )$ must be extracted by enforcing the relation:

\begin{equation}
    \frac{1}{2m} \sum_{i=1}^{2m} \delta(l_{i}^{+}, l^{+}) \delta(\gamma_{i}, \gamma) \, x_i = Q(\gamma) B(l^{+} \mid \gamma) \, x_0
\end{equation}

But we can write the left-hand side as:

\begin{eqnarray}
 \frac{1}{2m} \sum_{i=1}^{2m} \delta(l_{i}^{+}, l^{+}) \delta(\gamma_{i}, \gamma) \, x_i \!\!\!\! &=& \!\!\!\!  \frac{1}{2m} \sum_{k=1}^{m/2} \delta(l_{2k-1}^{+}, l^{+}) \delta(\gamma_{2k-1}, \gamma) \, x_{2k-1}+\frac{1}{2m} \sum_{i=1}^{m/2} \delta(l_{2k}^{+}, l^{+}) \delta(\gamma_{2k}, \gamma) \, x_{2k} \nonumber \\ 
 \!\!\!\! &\sim& \!\!\!\! \frac{x_0}{2} Q(\gamma) \Big\{ B(l^{+} \mid \gamma) + B(\gamma - l^{+}-1 \mid \gamma) \, (1-\delta(l^{+},\gamma))\, w(\gamma - l^{+}-1 \mid \gamma) + \nonumber\\
& & \:\:\:\:\:\:\:\:\:\:\:\:\:\:\:\:\:\:\:\:\:\:\:\:\:\:\:\: + B(\gamma \mid \gamma) \delta(l^{+},\gamma) w(\gamma \mid \gamma) \Big\}
\end{eqnarray}

The last equality is valid only when $N \gg 1$. It can be satisfied only if:

\begin{eqnarray}
    w(\gamma \mid \gamma)\!\!\!\! &=& \!\!\!\! 1 \\
    w(\gamma - l^{+}-1 \mid \gamma) \!\!\!\! &=& \!\!\!\! \frac{B(l^{+} \mid \gamma)}{B(\gamma - l^{+}-1 \mid \gamma)} = \frac{l^{+}}{\gamma - l^{+}} \, , \:\:\:\:\:\:\: \text{if} \:\:0 \leq l^{+} < \gamma
\end{eqnarray}

The assignment of $x_{2k}$ can be then synthesized into the expression:

\begin{eqnarray}
    x_{2k} = \frac{l_{2k}^{+}+1}{l_{2k-1}^{+}+1} x_{2k-1}
\end{eqnarray}

Given that $x_{2k-1} = x_0$ and $l_{2k-1}^{\pm}$, $l_{2k}^{\pm}$ are chosen as we explained above. To extend this to the case of the population of probabilities $P(\vec{s}; \vec{l}^{+}, \vec{l}^{-})$, we just need to apply the same rules to each pair $(l_z^{+}, l_z^{-})$, taking into account that the object $x_0$ is now a probability $p_0(\vec{s})$ defined over all the configurations of the vector $\vec{s}$. 

\begin{figure}%[tbhp]
\centering  

\subfloat[]{
\includegraphics[width=.43\linewidth]{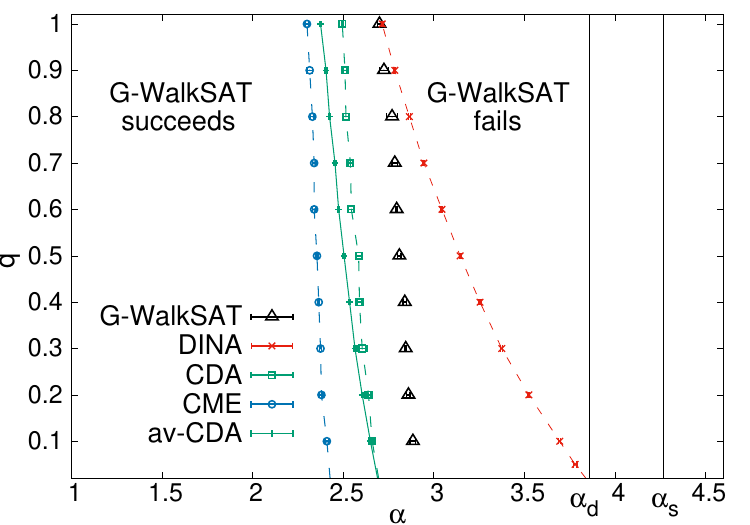}
}
\subfloat[]{
\includegraphics[width=.43\linewidth]{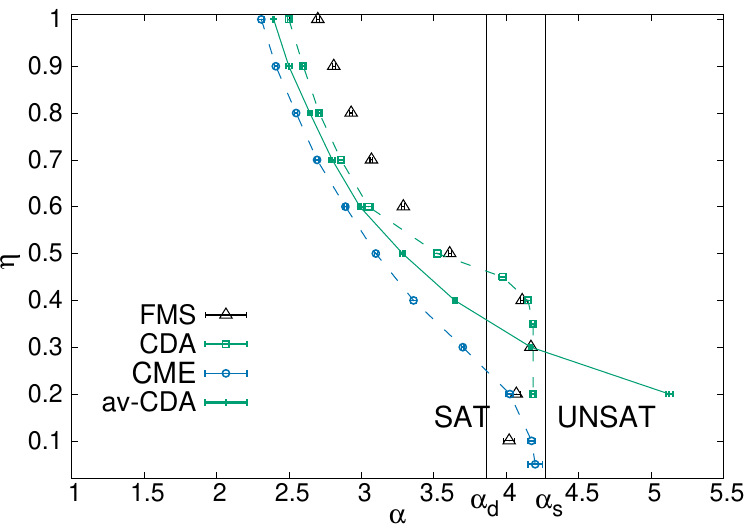}
}
    \caption{Phase diagrams for the algorithms (triangles) in the random 3-SAT, together with predictions of DINA, CME, CDA, and the average case version of the CDA (av-CDA). Both panels show the thresholds for several values of the algorithmic parameter ($q$ for G-WalkSAT and $\eta$ for FMS). To obtain them, we plot energy \textit{vs.} time and observe the curvature for long times. The transition is defined at the $\alpha$ where the type of curvature changes. Vertical lines mark the dynamical transition $\alpha_d \approx 3.86$ and the SAT-UNSAT transition $\alpha_s \approx 4.267$. \textbf{a)} G-WalkSAT was run in formulas with $N=5 \times 10^{4}$ variables. For the CDA and the CME, the system size was $N=10^{5}$ for $q \geq 0.2$ and $N=5 \times 10^{4}$ for $q\leq0.1$. \textbf{b)} FMS was run in formulas with $N=5 \times 10^{4}$ variables for $\eta \geq 0.6$, and with $N=\times 10^{5}$ variables for $\eta \leq 0.5$. For the CDA and the CME we used $N=5 \times 10^{4}$, except for the last two points of the CME that were obtained with $N=5 \times 10^{5}$.}
    \label{fig:phase_diagrams_av_CDA1}
\end{figure}

With this scheme, we can run the average case version of the CDA without too much computational effort. The results for Focused Metropolis Search and G-WalkSAT are presented in Fig. \ref{fig:phase_diagrams_av_CDA1}, where they are labeled as \textit{av-CDA}. The predictions of this av-CDA work well provided that the underlying algorithmic dynamics are far from the zone $\alpha_d < \alpha < \alpha_s$. This includes all values of $q$ for G-WalkSAT and the range $\eta > 0.5$ for FMS. 

Below $\eta=0.5$, the average case predictions for FMS's dynamics are completely wrong, indicating that a correct description should include more information about the local structure of the formulas. Indeed, the single instance version of the CDA is instead capable of predicting the algorithmic threshold (see Fig. 3 in the main text). More than that, it captures the local structures of the solutions (see Fig. 4 in the main text and Fig. \ref{fig:CDA-d_several_tau} here). Thus, one would need to add this local information to a proper average case theory, which is not a simple task.

\section{Transition rules for G-WalkSAT} \label{sec:trans_GW}

As we pointed out in the main text, one needs to encode the algorithmic dynamics into rules $r_i(\sigma_i, \sigma_{\partial i})$. The key point with G-WalkSAT is to write the probability of flipping $\sigma_i$ in a greedy step, given it belongs to $S$ satisfied clauses. Figs. 2 and 3 in the main text are produced assuming that all the neighbors of $\sigma_i$ have the same probabilities $p(S'=S)$ and $p(S' > S)$. The first is the probability that a neighbor belongs to the same number $S'=S$ of satisfied clauses, and the second is the probability of having $S'>S$. Averaging over the neighbor's connectivity, we obtain:

% \begin{eqnarray}
%     p(S'=S) \!\!\!\! &=&\!\!\!\! \sum_{\gamma=S}^{\infty} e^{-\alpha K} \frac{(\alpha K)^{\gamma}}{\gamma!} \, \binom{\gamma}{S} \, \big(p_{sat} \big)^{S} \, \big( 1 - p_{sat} \big)^{\gamma - S} = \Big(\frac{p_{sat}}{1 - p_{sat}}\Big)^{S} \frac{e^{\alpha K}}{S!} \sum_{\gamma=S}^{\infty} \frac{\big[\alpha K (1 - p_{sat})\big]^{\gamma}}{(\gamma - S)!} \nonumber \\
%     p(S'=S) \!\!\!\! &=&\!\!\!\! \frac{\big[\alpha K p_{sat}\big]^{S}}{S!} e^{-\alpha K p_{sat}} \label{eq:computing_P_S_av}
% \end{eqnarray}
\begin{equation}
    p(S'=S) = \sum_{\gamma=S}^{\infty} e^{-\alpha K} \frac{(\alpha K)^{\gamma}}{\gamma!} \, \binom{\gamma}{S} \, \big(p_{sat} \big)^{S} \, \big( 1 - p_{sat} \big)^{\gamma - S}  = \frac{\big[\alpha K p_{sat}\big]^{S}}{S!} e^{-\alpha K p_{sat}} \label{eq:computing_P_S_av}
\end{equation}
where $\gamma$ is the connectivity of the neighbor minus one, which in Poisson random graphs also follows the Poisson distribution with the same mean $\alpha K$.

We have thus obtained that $p(S'=S)$ is the Poisson distribution with mean $\alpha K p_{sat}$, where $p_{sat}$ is the probability of finding a satisfied clause in the neighborhood of a variable. We can also assume $p_{sat}$ to be independent on the site and substitute it by $p_{sat} = 1 - \langle E \rangle / M$. The probability $p(S'>S)$ is given by the sum:

\begin{equation}
    p(S'>S) = \sum_{S'=S+1}^{\infty} \frac{\big[\alpha K p_{sat}\big]^{S'}}{(S')!} e^{-\alpha K p_{sat}}
\end{equation}

\section{Completely greedy G-WalkSAT} \label{sec:GW_q0}

In Section \ref{sec:trans_GW} we gave closed forms for the transition rules used in Figs. 2 and 3 in the main text. The results show these expressions to be sufficient to describe most of the phase diagram of G-WalkSAT. However, when most steps are greedy ($q \leq 0.01$), some extra details are necessary to capture the dynamics with the rule $r_i(\sigma_i, \sigma_{\partial i})$. 

\begin{figure}%[tbhp]
\centering
  \subfloat[]{
	   \centering
	   \includegraphics[width=.43\linewidth]{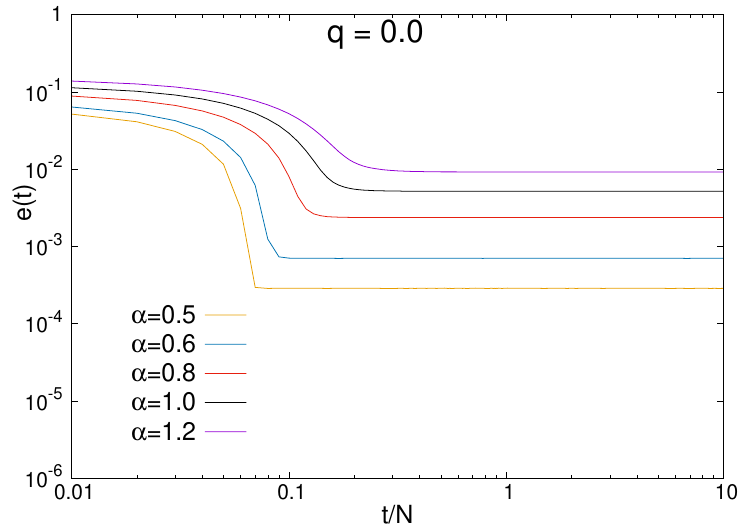} \label{fig:CDA_det_q0}}
            \subfloat[]{
	   \centering
	   \includegraphics[width=.43\linewidth]{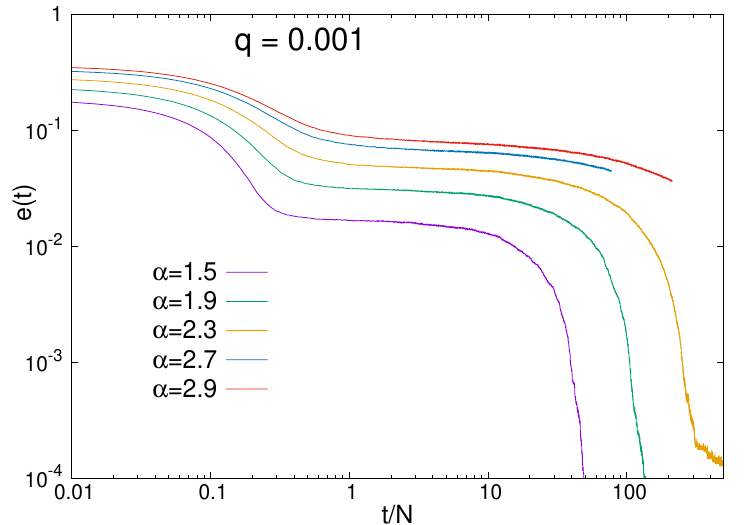} \label{fig:CDA_det_q001}}
	   
	   \subfloat[]{
	   \centering
	   \includegraphics[width=.43\linewidth]{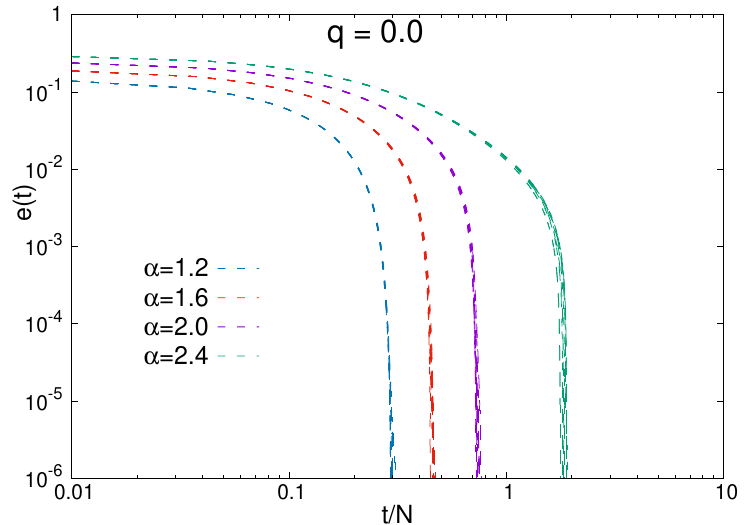}\label{fig:CDA_av_q0}}
            \subfloat[]{
	   \centering
	   \includegraphics[width=.43\linewidth]{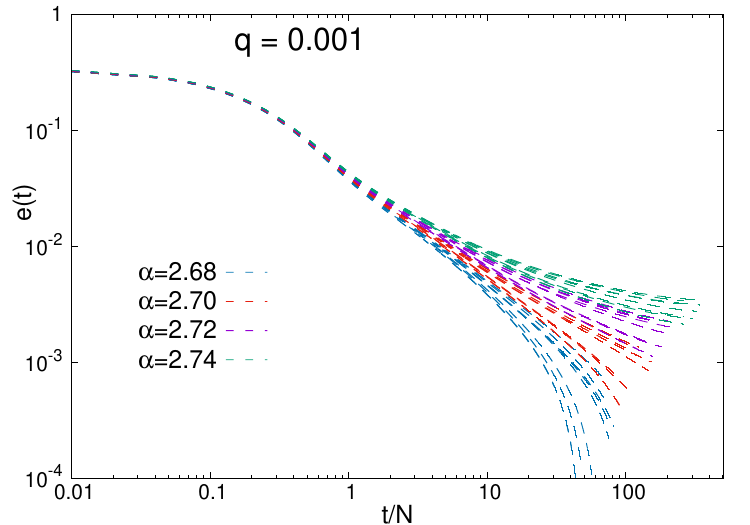}\label{fig:CDA_av_q001}}
	   
	   \subfloat[]{
	   \centering
	   \includegraphics[width=.43\linewidth]{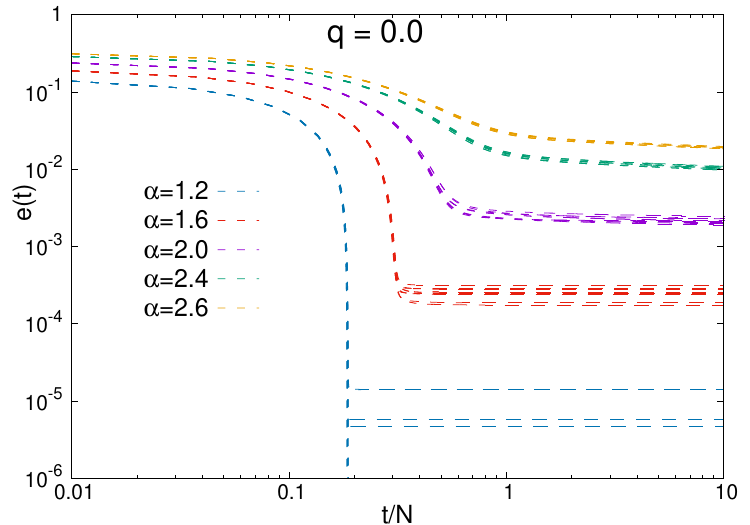} \label{fig:GW_q0}}
            \subfloat[]{
	   \centering
	   \includegraphics[width=.43\linewidth]{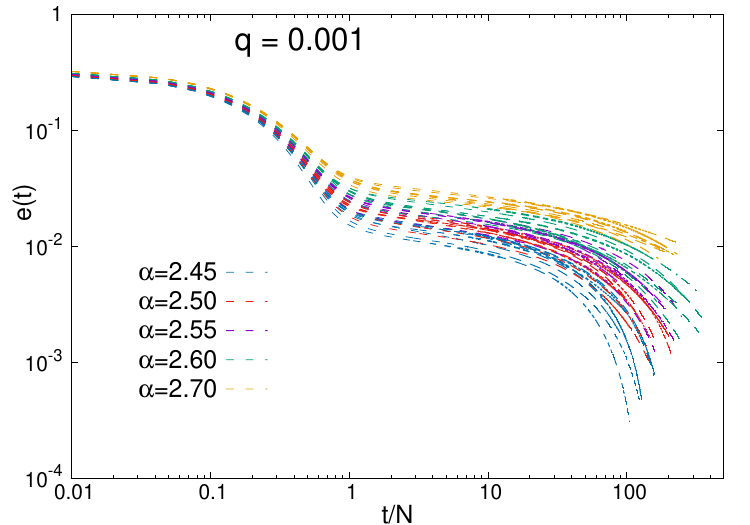} \label{fig:GW_q001}}
	   
    \caption{Algorithmic dynamics of G-WalkSAT in the random 3-SAT. All panels show the time evolution of the energy density for different values of $\alpha$ in logarithmic scale. The variables are initially assigned to be $0$ or $1$ with the same probability. Panels {\bf a)} and  {\bf b)} show the average of G-WalkSAT results over $1000$ instances for $q=0$ and $q=0.001$. System size is $N=5 \times 10^{4}$. Panels {\bf c)} and  {\bf d)} show the results of the simplest version of CDA applied to $8$ different formulas with algorithmic parameter $q=0$ and $q=0.001$. System size is $N=5 \times 10^{4}$. Panels {\bf e)} and  {\bf f)} show the results of CDA with the modified rates. The equations were run in $8$ different formulas with algorithmic parameter $q=0$ and $q=0.001$. System sizes are $N=5 \times 10^{3}$ (panel {\bf e)}) and $N=10^{3}$ (panel {\bf f)})}
    \label{fig:GWALK_low_q}
\end{figure}

The behavior of G-WalkSAT close to $q=0$ is pathological, as can be seen in Figs. \ref{fig:GW_q0} and \ref{fig:GW_q001}. Exactly at $q=0$, the algorithm fails to solve the formulas for any possible $\alpha$, remaining blocked at very low energies. At $q=0.001$, it shows a long plateau after which the energy decreases to small values. However, for $\alpha$ high enough the algorithm remains again blocked at those small energies, announcing what will be the behavior at $q=0$. 

The simple versions of the CDA in the main text cannot reproduce it (Eqs. (7) and (8)). At $q=0.001$, the curves in Fig. \ref{fig:CDA_av_q001} do not show any plateau and indicate a fast convergence to zero up to $\alpha=2.68$, but the algorithm is having problems in finding solutions already at $\alpha=2.3$ (see Fig. \ref{fig:GW_q001}). When all the steps are greedy ($q=0$), the results in Fig. \ref{fig:CDA_av_q0} predict polynomial time solutions at least up to $\alpha=2.4$. We know that, when the system size is large enough, G-WalkSAT cannot solve instances even at $q=0.5$.

This is not the end. We can insert some information into the probability $p(S'=S)$ that will make it site-dependent. Conditioning one neighbor $j \in \partial i$ to have connectivity $c_j$, with $c_j \geq 1$, we get simple forms for both probabilities:

\begin{eqnarray}
    p(S_j \mid c_j) \!\!\!\! &=&\!\!\!\! \binom{c_j - 1}{S_j} \big( p_{sat}\big)^{S_j} \, \big(1 - p_{sat} \big)^{c_j - 1 - S_j} \label{eq:p_Sj_GW_det}% \\
    %p(S'>S \mid c_j) \!\!\!\! &=&\!\!\!\! \sum_{S'=S+1}^{c_j-1} \binom{c_j - 1}{S'} \big( p_{sat}\big)^{S'} \, \big(1 - p_{sat} \big)^{c_j - 1 - S'}
\end{eqnarray}

This is the probability of finding a neighbor belonging to $S_j$ satisfied clauses. Now, the total rate (see Eq. (7) in the main text) has a more complicated expression:

\begin{equation}
 r_i^{\text{GW-CDA}}(\sigma_i, \sigma_{\partial i}) = (1 - q) \, \frac{E_i}{K\langle E \rangle} + \frac{q}{\langle E \rangle} \sum_{a \in \partial i} p_a(\text{g} | S, \vec{c}_{j \in a \setminus i}) \label{eq:rate_GW_CDA_det}
\end{equation}

The new probability of flipping $\sigma_i$ after selecting clause $a \in \partial i$ depends on the specific clause $a$ through the connectivities of the other variables inside $a$. These are inside the vector $\vec{c}_{j \in a \setminus i}$, and we have:

\begin{eqnarray}
    p_a(\text{g} | S, \vec{c}_{j \in a \setminus i}) = \Big[ \prod_{j \in a \setminus i} \sum_{S_j=S}^{c_j-1} \Big] \frac{1}{1+\sum_{j \in a \setminus i }\delta(S_j, S)} \prod_{j \in a \setminus i} p(S_j \mid c_j) \label{eq:pa_GW_det}
\end{eqnarray}
where the symbol $\Big[ \prod_{j \in a \setminus i} \sum_{S_j=S}^{c_j-1} \Big] $ represents $K-1$ sums, one for each of the numbers $S_j$, and $\delta(S_j, S)$ is a Kronecker delta.

Figs. \ref{fig:CDA_det_q0} and \ref{fig:CDA_det_q001} are produced with the more detailed transition rules in Eqs. \ref{eq:p_Sj_GW_det}, \ref{eq:rate_GW_CDA_det}, and \ref{eq:pa_GW_det}. The curves stop showing a fast convergence to zero. At $q=0.001$, the new CDA shows an initial plateau whose time scale is close to what we observe in the real algorithmic dynamics. The behavior at $q=0$ is also similar to what we get from G-WalkSAT. The new CDA predicts the dynamics to remain blocked at non-zero energies even at very small $\alpha$ (like $\alpha=1.2$ in Fig. \ref{fig:CDA_det_q0}).

The results in this section indicate that small improvements in the formulation of the transition rules $r_i$ can lead to important qualitative changes. Remarkably, including information about the connectivity of each site is enough to capture the right qualitative behavior of G-WalkSAT close to $q=0$.

\section{Other plots of the dynamics} \label{app:dyn}

As an example, the main text includes two graphics for the dynamics of G-WalkSAT and Focused Metropolis Search (FMS). Those were done for specific values of the algorithmic parameters $q=0.5$ and $\eta=0.5$, respectively. When we vary $\eta$ in the FMS, the time dependence of the energy displays some qualitative changes that merit further discussion.

\begin{figure}%[tbhp]
\centering
  \subfloat[]{
	   \centering
	   \includegraphics[width=.43\linewidth]{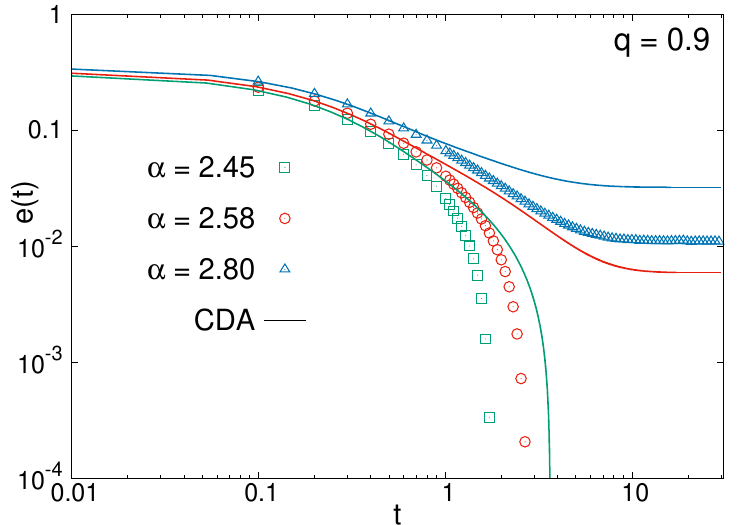}}
	   \subfloat[]{
	   \centering
	   \includegraphics[width=.43\linewidth]{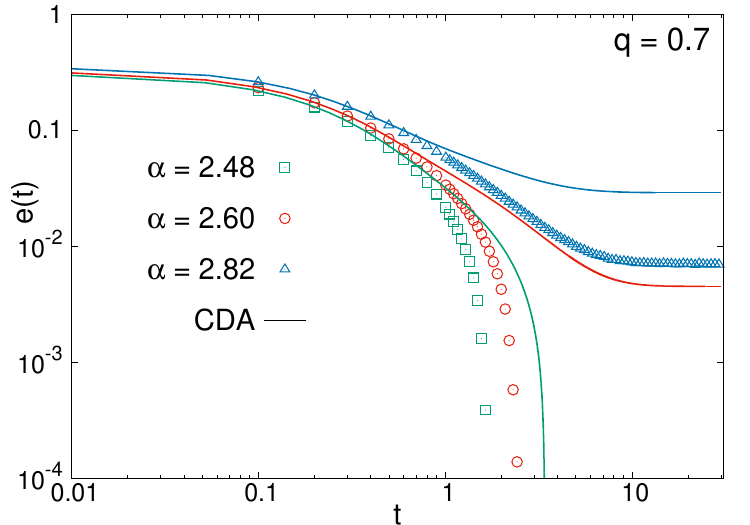}}
	   
	   \subfloat[]{
	   \centering
	   \includegraphics[width=.43\linewidth]{Figures/CDA_vs_WalkSAT_energy_q_50_paper.pdf}}
	   \subfloat[]{
	   \centering
	   \includegraphics[width=.43\linewidth]{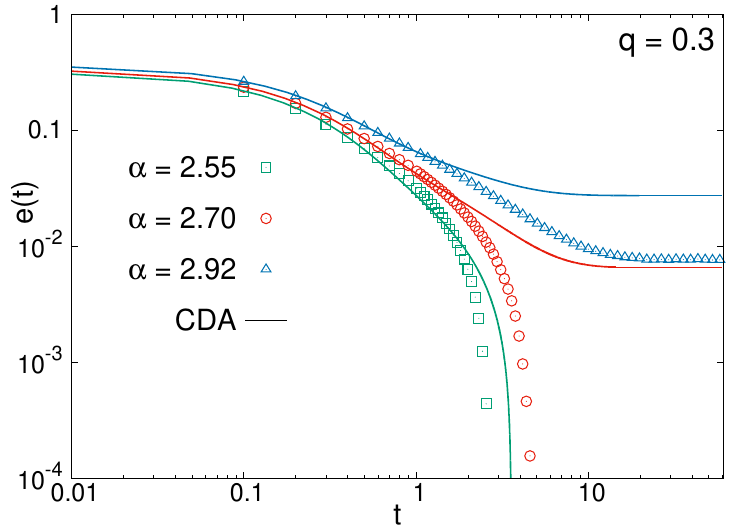}}
	   
    \caption{Algorithmic dynamics of G-WalkSAT in the random 3-SAT. All panels show the time evolution of the energy density for different values of $\alpha$ in logarithmic scale. The variables are initially assigned to be $0$ or $1$ with the same probability. Points represent an average over $s$ runs of the algorithm for a single 3-SAT formula. Lines are the prediction of the CDA for the algorithmic dynamics on the same formulas. System size is $N=5 \times 10^{4}$ in all cases. {\bf a)} $q=0.9$,  {\bf c)} $q=0.7$, {\bf e)} $q=0.9$, and {\bf g)} $q=0.3$}
    \label{fig:alg_dyn_GWALK}
\end{figure}

Fig. \ref{fig:alg_dyn_GWALK} shows that this is not the case for G-WalkSAT. The four graphics for $q=0.9, 0.7, 0.5, 0.3$ are very similar. The time scale to reach a solution is always less than $10$ steps (remember that each step involves $N$ iterations). Furthermore, the value of the energy density at long times when the algorithm does not find a solution is also of the same order (around $10^{-2}$). We checked that this behavior remains the same throughout the interval $q \in [0.01, 1]$. The dynamics at low $q$ will be treated in more detail in Section \ref{sec:GW_q0}.

\begin{figure}%[tbhp]
\centering
  \subfloat[]{
	   \centering
	   \includegraphics[width=.43\linewidth]{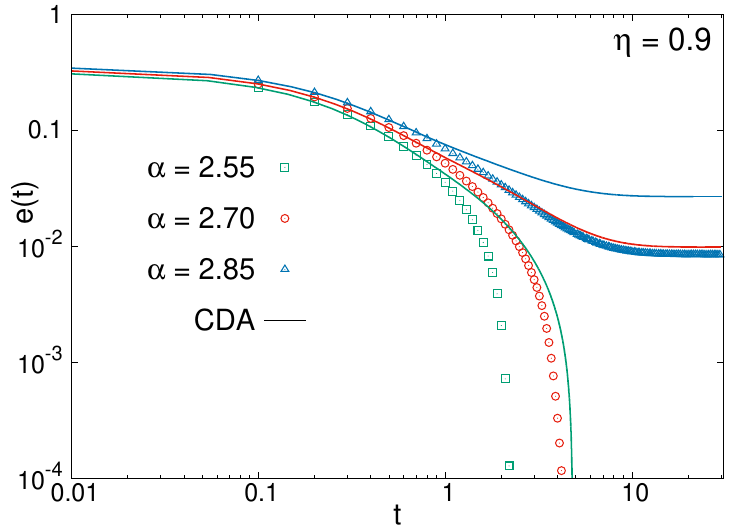}}
	   \subfloat[]{
	   \centering
	   \includegraphics[width=.43\linewidth]{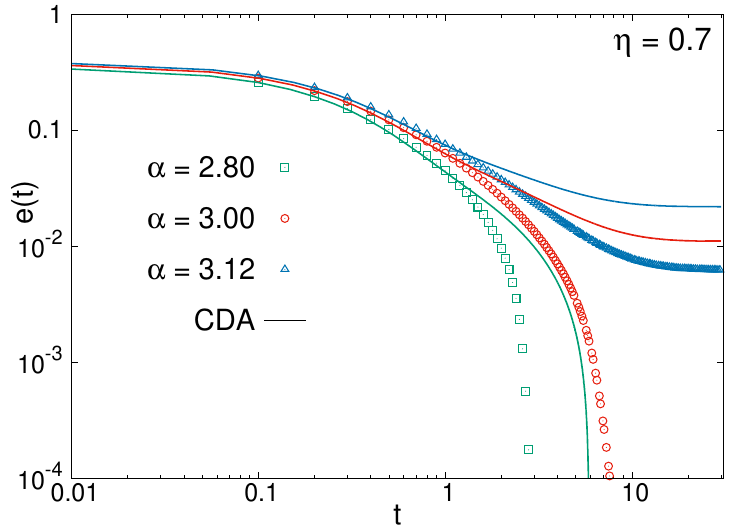}}
	   
	  \subfloat[]{
	   \centering
	   \includegraphics[width=0.43\linewidth]{Figures/CDA_vs_FMS_energy_eta_50_paper.pdf}}
	   \subfloat[]{
	   \centering
	   \includegraphics[width=.43\linewidth]{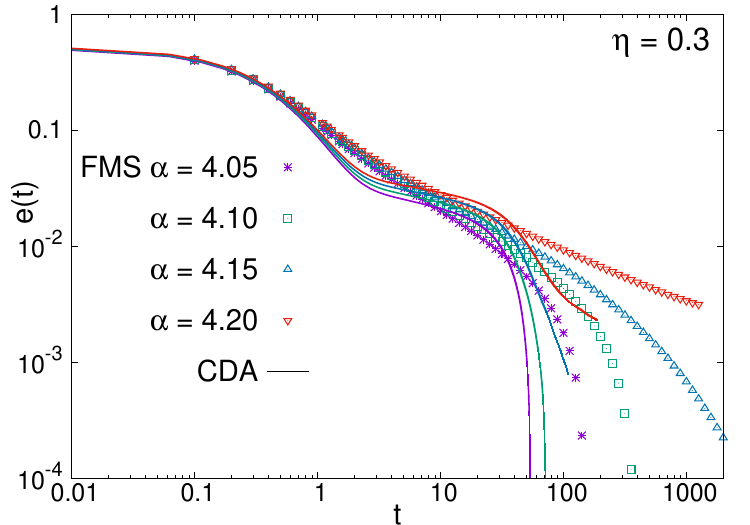}}
	   
    \caption{Algorithmic dynamics of FMS in the random 3-SAT. All panels show the time evolution of the energy density for different values of $\alpha$ in logarithmic scale. The variables are initially assigned to be $0$ or $1$ with the same probability. Points represent an average over $s$ runs of the algorithm for a single 3-SAT formula. Lines are the prediction of the CDA for the algorithmic dynamics on the same formulas. System size is $N=5 \times 10^{4}$ for $\eta \geq 0.5$ and $N=5 \times 10^{5}$ for $\eta=0.3$. {\bf a)} $\eta=0.9$,  {\bf c)} $\eta=0.7$, {\bf e)} $\eta=0.9$, and {\bf g)} $\eta=0.3$}
    \label{fig:alg_dyn_FMS}
\end{figure}

On the other hand, FMS's dynamics increase their characteristic time scale when the noise $\eta$ diminishes (see Fig. \ref{fig:alg_dyn_FMS}). From less than $10$ steps to get a solution at high $\eta$ ($0.9$ and $0.7$), the algorithm needs a bit more than $100$ steps at $\eta=0.5$. This change coincides with the proximity of the transition to the 'hard' phase ($\alpha_d < \alpha < \alpha_s$). Once inside that region, the time to solution spikes to more than $1000$ steps, as can be seen in the last panel of Fig. \ref{fig:alg_dyn_FMS} for $\eta=0.3$. 

The energy density close to the transition is also around $10^{-2}$ for high noise $\eta$. For $\eta=0.3$, FMS reaches low energies even when it does not rapidly converge to zero. Its decay is much smoother, signaling an enhancement of the criticality inside the 'hard' region.

All these features are qualitatively predicted by the CDA, which describes the dynamics taking into account only local correlations. Precisely this capacity to sense the presence of the 'hard' phase was exploited to convert the CDA into the new decimated algorithm we introduce in the main text of this article.

\section{Dynamic Independent Neighbor Approximation (DINA)} \label{sec:DINA_KSAT}

In the main text, we compared CDA's results with a technique from the literature: the Dynamic Independent Neighbor Approximation (DINA) \cite{BarthelwalkSAT2003}. The authors of Ref. \cite{BarthelwalkSAT2003} write closed differential equations for the probabilities $\hat{P}_t(u, s)$ of finding a variable in exactly $s$ satisfied clauses and $u$ unsatisfied clauses:

\begin{eqnarray}
\frac{d \hat{P}_t(u, s)}{dt} \!\!\!\! &=& \!\!\!\! -r(u, s) \, \hat{P}_t(u, s) + \nonumber \\
& & \!\!\!\!\!\!\!\!\!\!\!\!\!\!\!\!\!\!\!\!\!\!\!\!\!\!\!\! + \Big(\frac{1}{2^{K} - 1} \Big)^{u} \sum_{k = 0}^{s} \binom{u + k}{k} \Big(1 - \frac{1}{2^{K} - 1} \Big)^{k} \, r(s - k, u + k) \,  \hat{P}_t(s - k, u + k) - \nonumber \\
& & \!\!\!\!\!\!\!\!\!\!\!\!\!\!\!\!\!\!\!\!\!\!\!\!\!\!\!\! - \frac{(K - 1)}{2^{K} - 1} \frac{\langle s \, r(u, s) \rangle_{t}}{\langle s \rangle_{t}} \Big[ s \hat{P}_t(u, s) - (s + 1)\hat{P}_t(u-1, s+1) \Big] -
\label{eq:DINA_KSAT} \\
& & \!\!\!\!\!\!\!\!\!\!\!\!\!\!\!\!\!\!\!\!\!\!\!\!\!\!\!\! - (K - 1) \frac{\langle u \, r(u, s) \rangle_{t}}{\langle u \rangle_{t}} \Big[ u \hat{P}_t(u, s) - (u + 1) \hat{P}_t(u+1 , s-1) \Big], \nonumber
\end{eqnarray}
where $r(u, s)$ is the probability per time unit of flipping a variable that is in $s$ satisfied and $u$ unsatisfied clauses. Additionally, the notation $\langle \cdot \rangle$ denotes the average of any quantity weighted by the probabilities $\hat{P}_t(u, s)$. This means that $\langle \cdot \rangle \equiv \sum_{u} \sum_{s} \, [ \, \cdot \, ] \hat{P}_t(u, s)$.

Note that $\hat{P}_t(u, s)$ is the joint probability of a central variable $\sigma$ and all its neighbors. There are as many values of $\hat{P}^{t}(u, s)$ as there are combinations of the quantities $(u, s)$ of unsatisfied and satisfied clauses in that set. The first term on the right-hand side of Eq. \ref{eq:DINA_KSAT} represents the probability of flipping that central variable ($\sigma \to -\sigma$) and contributes negatively to the derivative of $\hat{P}^{t}(u, s)$.

The second term accounts for all positive contributions to the derivative of $\hat{P}^{t}(u, s)$ related to transitions $-\sigma \to \sigma$. When $\sigma$ is flipped, all $u$ unsatisfied clauses it belongs to become satisfied. Additionally, a number $k \leq s$ of the remaining clauses will remain satisfied. Only $s - k$ of them will become unsatisfied after the flip. Therefore, the second line of Eq. \ref{eq:DINA_KSAT} includes the probabilities $\hat{P}_t(s - k, u + k)$ of having exactly $s - k$ unsatisfied and $u + k$ satisfied clauses. These are multiplied by the binomial probability $p(k)=\binom{u + k}{k} \left(\frac{1}{2^{K} - 1}\right)^{u} \left(1 - \frac{1}{2^{K} - 1}\right)^{k}$ and summed over all possible values of $k$. Note that $2^{K} - 1$ is the number of configurations that satisfy a clause.

From the above description, it follows that DINA assumes all configurations satisfying a clause to be equiprobable, which is not true in general. Furthermore, to write Eq. \ref{eq:DINA_KSAT}, the authors of Ref. \cite{BarthelwalkSAT2003} make a factorization of a joint probability distribution that is similar in spirit to what we did to obtain CDA. Indeed, the follow-up work of Ref. \cite{Semerjiandyn2004} clarifies that Eq. \ref{eq:DINA_KSAT} is obtained by assuming the relations:

\begin{eqnarray}
\hat{P}_{t}^{S}(\hat{u}, \hat{s}, u, s) \!\!\!\! &\approx& \!\!\!\!   \frac{\hat{s} \, \hat{P}_{t}(\hat{u} , \hat{s})}{\sum_{s'} \sum_{u'} s' \, \hat{P}_{t}(u', s')} \, P_t(u, s) \, \mathbb{I}(s>0) = \frac{\hat{s} \, \hat{P}_{t}(\hat{u} , \hat{s})}{\langle s \rangle_t} \, P_t(u, s) \, \mathbb{I}(s>0) \label{eq:closure_DINA_SAT} \\
\hat{P}_{t}^{U}(\hat{u}, \hat{s}, u, s) \!\!\!\! &\approx& \!\!\!\!  \frac{\hat{u} \, \hat{P}_{t}(\hat{u} , \hat{s})}{\sum_{s'} \sum_{u'} u' \, \hat{P}_{t}(u', s')}  \, P_t(u, s)  \, \mathbb{I}(u>0) = \frac{\hat{u} \, \hat{P}_{t}(\hat{u} , \hat{s})}{\langle u \rangle_t}  \, P_t(u, s)  \, \mathbb{I}(u>0)\label{eq:closure_DINA_UNSAT}
\end{eqnarray}
where $\mathbb{I}(x>0)$ is an indicator function that evaluates to one if $x>0$, and to zero otherwise. In Eq. \ref{eq:closure_DINA_SAT}, $P_t^{S}(\hat{u}, \hat{s}, u, s)$ is the probability to find a variable belonging to $\hat{u}$ unsatisfied and $\hat{s}$ satisfied clauses, another variable in $u$ unsatisfied and $s$ satisfied clauses, and that they share one of those satisfied clauses. In other words, one of the $\hat{s}$ satisfied clauses is also one of the $s$ satisfied clauses. The key point in DINA is to factorize this joint probability so that everything is expressed in terms of $P_t(u,s)$, as shown in Eq. \ref{eq:closure_DINA_SAT}. Something analogous happens with $P_t^{U}(\hat{u}, \hat{s}, u, s)$ in Eq. \ref{eq:closure_DINA_UNSAT}, which is the joint probability in the case that the two variables share one unsatisfied clause.

The approximations in Eqs. \ref{eq:closure_DINA_SAT} and \ref{eq:closure_DINA_UNSAT} are fundamental to obtain the third and fourth lines of Eq. \ref{eq:DINA_KSAT}. These contain contributions to the derivative due to flips of the neighbors of the central variable.

If the rules $r(u, s)$ are known, Eq. \ref{eq:DINA_KSAT} can be numerically integrated to obtain the temporal dependence of $\hat{P}_{t}(u, s)$ and with it a prediction for the evolution of the number of unsatisfied clauses per variable $e(t) = \langle u \rangle_{t} / K$. This is done in Ref. \cite{BarthelwalkSAT2003} to study the dynamics of G-WalkSAT, whose dynamic rules are functions $r(u, s)$ of $u$ and $s$. We complement their results by obtaining DINA's prediction of G-WalkSAT's phase diagram, presented in the upper panel of Fig. 3 in the main text. The initial conditions are chosen as follows:

\begin{equation}
 \hat{P}^{0}(u, s) = e^{-K \alpha } \frac{(K \alpha)^{u + s}}{u! s!} \, \Big(1 - \frac{1}{2^{K}} \Big)^{s} \, \Big(\frac{1}{2^{K}} \Big)^{u}.
\end{equation}

On the other hand, FMS has dynamic rules that cannot be written in the form $r(u,s)$. With this algorithm, it is important to know how many clauses $u^{+}$ will become unsatisfied after flipping the variable or, equivalently, the number $k$ of clauses that will remain satisfied. The dynamic rule is a function $r_{\text{FMS}}(u, k)$. However, if we do not want to go beyond DINA's approximations, we have the relation: \(\hat{P}(u, k, s) \equiv \hat{P}(u, s) \, \binom{s}{k} \big[1 / (2^{K}-1)\big]^{s - k} \,\big[1 - 1 / (2^{K}-1)\big]^{k}\). We can insert this back into Eq. \ref{eq:DINA_KSAT} to get:

\begin{eqnarray}
\frac{d \hat{P}_t(u, s)}{dt} \!\!\!\! &=& \!\!\!\! -r_{\text{eff}}(u,s) \, \hat{P}_t(u, s) + \nonumber \\
& & \!\!\!\!\!\!\!\!\!\!\!\!\!\!\!\!\!\!\!\!\!\!\!\!\!\!\!\! + \Big(\frac{1}{2^{K} - 1} \Big)^{u} \sum_{k = 0}^{s} \binom{u + k}{k} \Big(1 - \frac{1}{2^{K} - 1} \Big)^{k} \, r_{\text{FMS}}(s - k, k) \,  \hat{P}_t(s - k, u + k) - \nonumber \\
& & \!\!\!\!\!\!\!\!\!\!\!\!\!\!\!\!\!\!\!\!\!\!\!\!\!\!\!\! - \frac{(K - 1)}{2^{K} - 1} \frac{\langle s \, r_{\text{eff}}(u, s) \rangle_{t}}{\langle s \rangle_{t}} \Big[ s \hat{P}_t(u, s) - (s + 1)\hat{P}_t(u-1, s+1) \Big] -
\label{eq:DINA_KSAT_FMS} \\
& & \!\!\!\!\!\!\!\!\!\!\!\!\!\!\!\!\!\!\!\!\!\!\!\!\!\!\!\! - (K - 1) \frac{\langle u \, r_{\text{eff}}(u, s) \rangle_{t}}{\langle u \rangle_{t}} \Big[ u \hat{P}_t(u, s) - (u + 1) \hat{P}_t(u+1 , s-1) \Big], \nonumber
\end{eqnarray}
where we introduced the effective transition rule:

\begin{equation}
    r_{\text{eff}}(u,s) = \sum_{k=0}^{s}  \, \binom{s}{k} \, \Big(\frac{1}{2^{K}-1} \Big)^{s - k} \,\Big(1 - \frac{1}{ 2^{K}-1} \Big)^{k} \, r_{FMS}(u, k) 
\end{equation}

Our adaptation of DINA to FMS allowed us to produce DINA's prediction of FMS's phase diagram, which the reader can see in the bottom panel of Fig. 3 in the main text.

\section{Improving the performance of CDA-guided decimation}

In the main text, we introduced a new decimation scheme to take advantage of the marginal predicted by the CDA. We showed that the algorithm solves 3-SAT instances beyond $\alpha_d=3.86$ using FMS's dynamic rules and a specific value of the parameter $\tau$ ($\tau=5$).

\begin{figure}%[tbhp]
\centering
  \subfloat[]{
	   \centering
	   \includegraphics[width=.43\linewidth]{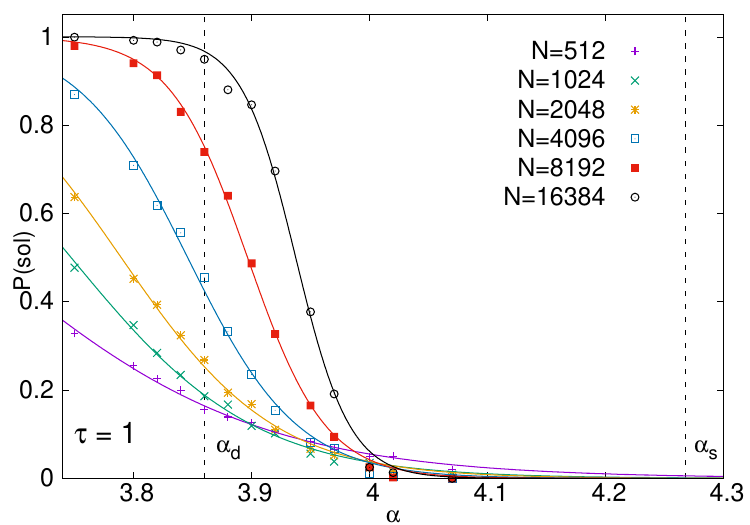} \label{fig:CDA-d_tau_1}}
	   
            \subfloat[]{
	   \centering
	   \includegraphics[width=.43\linewidth]{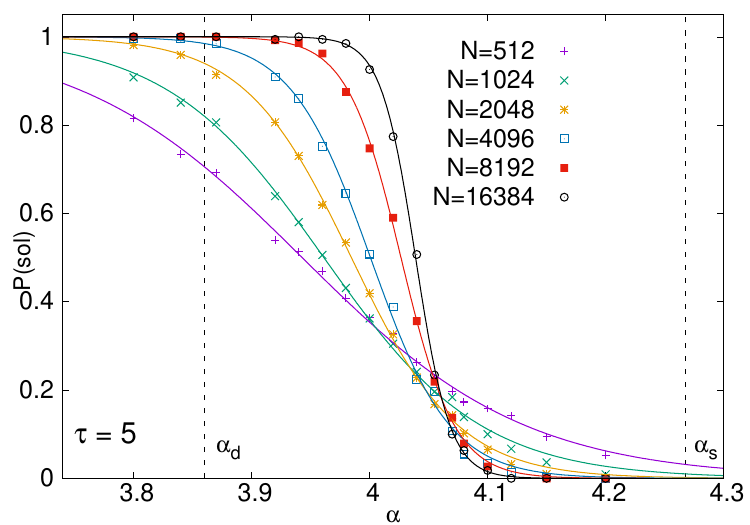} \label{fig:CDA-d_tau_5}}

           \subfloat[]{
	   \centering
	   \includegraphics[width=.43\linewidth]{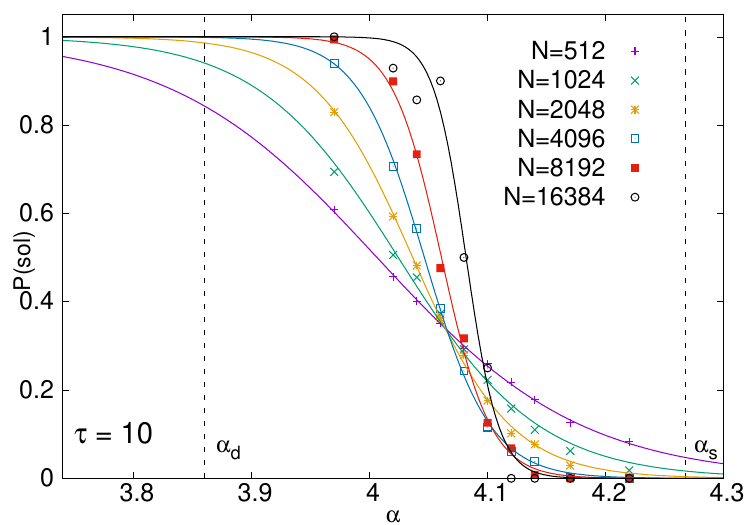} \label{fig:CDA-d_tau_10}}
    \caption{CDA-guided decimation with FMS rates in 3-SAT formulas for various system sizes $N$. Points represent the fraction of the instances solved for different values of $\alpha$. The variables were decimated one at a time, taking $\tau$ steps of the numeric integrator between consecutive reductions of the formula. The statistics include $1000$ formulas for each pair $(\alpha, N)$. Lines are logistic fits to the points. Panels {\bf (a)}, {\bf (b)}, and {\bf (c)} are produced with $\tau=1$, $\tau=5$, and $\tau=10$, respectively.}
    \label{fig:CDA-d_several_tau}
\end{figure}

The results in Fig. \ref{fig:CDA-d_several_tau} show that the performance of CDA-guided decimation improves when we increase $\tau$. The curves for the different system sizes move to the right, \textit{i.e.}, to larger values of $\alpha$, when $\tau$ increases from $\tau=1$ to $\tau=5$, and the same happens between $\tau=5$ and $\tau=10$.

\section{Efficient implementation of the CDA} \label{app:impl_CDA} 

As presented in the main text, the equation for the CDA
\begin{eqnarray}
\frac{dP_t(\vec{\sigma}_a)}{dt} = -\sum_{i \in a} \sum_{\vec{\sigma}_{\partial i \setminus a}} r_i(\sigma_i, \vec{\sigma}_{\partial i}) \, \Big[ \prod_{b \in \partial i \setminus a} \frac{P(\vec{\sigma}_{b})}{P(\sigma_i)} \Big] P(\vec{\sigma}_a) +  \sum_{i\in a} \sum_{\vec{\sigma}_{\partial i \setminus a}} r_i(-\sigma_i, \vec{\sigma}_{\partial i}) \, \Big[ \prod_{b \in \partial i \setminus a} \frac{P(F_{i}[\vec{\sigma}_{b}])}{P(-\sigma_i)} \Big] P(F_{i}[\vec{\sigma}_a])
 \label{eq:CDA1_app}
\end{eqnarray}
involves sums $\sum_{\vec{\sigma}_{\partial i \setminus a}}$ with a number of terms that grows exponentially with the connectivity of site $i$ in the graph. Indeed, the vector $\vec{\sigma}_{\partial i \setminus a}$ has $(c_i-1) (K-1)$ components, each one with two possible values. The sum has a number of terms scaling as $2^{K c_i}$.

To implement this sum more efficiently, we used the symmetries of the transition rules $r_i$ of the algorithms Focused Metropolis Search and G-WalkSAT. These are functions that depend, at most, on the number $u^{-}$ of unsatisfied clauses in the neighborhood of the variable and the number $u^{+}$ of clauses that will become unsatisfied after flipping the variable. More explicitly, we have:

\begin{eqnarray}
    r^{\text{FMS-CDA}}(u^{-}, u^{+}; \langle E \rangle) \!\!\!\! &=& \!\!\!\! \frac{u^{-}}{K \langle E \rangle} \text{min}\{1, \eta^{u^{+}-u^{-}} \} \label{eq:rate_FMS_av} \\
    r^{\text{GW-CDA}}(u^{-}, u^{+}; \langle E \rangle) \!\!\!\! &\equiv& \!\!\!\! r^{\text{GW-CDA}}(u^{-}; \langle E \rangle) = (1-q) \frac{u^{-}}{K \langle E \rangle} + q \frac{u^{-}}{\langle E \rangle} p(g \mid S) \label{eq:rate_GW_av}
\end{eqnarray}
where we used the probability $p(g \mid S)$ whose definition is given in the main text (see Eq. (8)).

Now, we can write:

\begin{equation}
 \sum_{\vec{\sigma}_{\partial i \setminus a}} r_i(\sigma_i, \vec{\sigma}_{\partial i}) \, \Big[ \prod_{b \in \partial i \setminus a} \frac{P(\vec{\sigma}_{b})}{P(\sigma_i)} \Big] = \sum_{u^{-}=0}^{l^{-}} \sum_{u^{+}=0}^{l^{+}} r\big[u^{-}+f(\vec{\sigma}_a), u^{+}+f(F_i[\vec{\sigma}_a]) \big] \!\!\!\! \sum_{\vec{\sigma}_{\partial i^{+}\setminus a} \mid u^{+}} \Big[ \prod_{b \in \partial i^{+} \setminus a} \frac{P(\vec{\sigma}_{b})}{P(\sigma_i)} \Big] \sum_{\vec{\sigma}_{\partial i^{-}\setminus a}  \mid u^{-}} \Big[ \prod_{b \in \partial i^{-} \setminus a} \frac{P(\vec{\sigma}_{b})}{P(\sigma_i)} \Big] \label{eq:sum_u_1}
\end{equation}
where $f(\vec{\sigma}_a) = 0$ if the configuration $\vec{\sigma}_a$ satisfies the clause, and $f(\vec{\sigma}_a) = 1$ otherwise. Remember that the operator $F_i[\vec{\sigma}_a]$ flips the value of $\sigma_i$ and leaves the rest of the vector untouched. The set $\partial i^{+} \setminus a$ contains the clauses in $\partial i \setminus a$ that are satisfied when the $i$-th variable takes the value $\sigma_i$. On the other hand, $\partial i^{+} \setminus a$ is formed by the clauses in $\partial i \setminus a$ that are satisfied when it takes the value $-\sigma_i$. The parameter $l^{+}$ is the number of clauses in $\partial i^{+} \setminus a$, and $l^{-}$ is the number of clauses in $\partial i^{-} \setminus a$. The sum $\sum_{\vec{\sigma}_{\partial i^{\pm}\setminus a} \mid u^{\pm}}$ is taken over all the configurations of $\vec{\sigma}_{\partial i^{\pm}\setminus a}$ compatible with $u^{\pm}$.

In Eq. \ref{eq:sum_u_1}, the sums $\sum_{\vec{\sigma}_{\partial i^{\pm}\setminus a} \mid u^{\pm}}$ are still in principle exponential in the number of neighbors. Notice, however, that if we find a way to compute these sums polynomially in $l^{\pm}$, then the whole operation becomes polynomial. Thus, the problem reduces to find a polynomial algorithm to compute the generic sum

\begin{equation}
    G(u; l, \sigma_i) = \sum_{\{\vec{\sigma}_{b \setminus i} \} \mid u}   \prod_{b} \frac{P(\vec{\sigma}_{b \setminus i}, \sigma_i)}{P(\sigma_i)}  = \sum_{\{s_{b \setminus i}\}} \delta\Big(\sum_{b} s_{b \setminus i},\, u \Big)  \prod_{b} \sum_{\vec{\sigma}_{b \setminus i} \mid s_{b \setminus i}} P(\vec{\sigma}_{b \setminus i} \mid \sigma_i)
\end{equation}
where $\delta(a,b)$ is the Kronecker delta. The new variable $s_{b \setminus i}=0$ if the variables in $b \setminus i$ satisfy the clause, $s_{b \setminus i}=1$ otherwise. The sum $\sum_{\vec{\sigma}_{b \setminus i} \mid s_{b \setminus i}}$ is taken over all configurations $\vec{\sigma}_{b \setminus i}$ compatible with $s_{b \setminus i}$. Introducing a conditional probability in the new variable $P(s_{b \setminus i} \mid \sigma_i) = \sum_{\vec{\sigma}_{b \setminus i} \mid s_{b \setminus i}} P(\vec{\sigma}_{b \setminus i} \mid \sigma_i)$, we get a simplified expression:

\begin{equation}
    G(u; l, \sigma_i) = \sum_{\{s_{b \setminus i}\}} \delta\Big(\sum_{b} s_{b \setminus i},\, u \Big)  \prod_{b} P(s_{b \setminus i} \mid \sigma_i)
\end{equation}

To compute all these sums, we use a recursive algorithm. We choose an arbitrary order for the clauses $b_1, b_2, \ldots, b_{l}$. Let us start by assigning $G(0; 0, \sigma_i) = 1$. Then, we impose the relations:

\begin{eqnarray}
    G(0; k+1, \sigma_i) \!\!\!\! &=& \!\!\!\! P(s_{b_k \setminus i} = 0 \mid \sigma_i) \, G(0; k, \sigma_i) \nonumber \\
    G(u; k+1, \sigma_i) \!\!\!\! &=& \!\!\!\! P(s_{b_k \setminus i} = 0 \mid \sigma_i) \, G(u; k, \sigma_i) + P(s_{b_k \setminus i} = 1 \mid \sigma_i) \, G(u-1; k, \sigma_i)\:\:\:\:\:\:\:\:\:\:\:\:\:\:\:\:0< u \leq k \nonumber \\
    G(k+1; k+1, \sigma_i) \!\!\!\! &=& \!\!\!\! P(s_{b_k \setminus i} = 1 \mid \sigma_i) \, G(u-1; k, \sigma_i)
\end{eqnarray}
from $k=0$ to $k=l-1$. This allows us to compute all $G(u; l, \sigma_i)$, with $u=0, \ldots, l$, with a number of operation that is proportional to $l^{2}$. We save all those values and insert them back into Eq. \ref{eq:sum_u_1} to get:

\begin{equation}
 \sum_{\vec{\sigma}_{\partial i \setminus a}} r_i(\sigma_i, \vec{\sigma}_{\partial i}) \, \Big[ \prod_{b \in \partial i \setminus a} \frac{P(\vec{\sigma}_{b})}{P(\sigma_i)} \Big] = \sum_{u^{-}=0}^{l^{-}} \sum_{u^{+}=0}^{l^{+}} r\big[u^{-}+f(\vec{\sigma}_a), u^{+}+f(F_i[\vec{\sigma}_a]) \big] \, G(u^{+}; l^{+}, \sigma_i)\, G(u^{-}; l^{-}, \sigma_i) \label{eq:sum_u_2}
\end{equation}

The new sum also takes $O(l^{2})$ operations. Remembering that, in average, $l^{\pm}$ is proportional to the connectivity $c$, we realize we went from a exponential computational cost ($O(2^{Kc_i})$) to a polynomial cost $O(c_i^{2})$. This allows us to run the CDA much faster and to reach system sizes $N \sim 10^{5}$ close to the SAT/UNSAT transition in random 3-SAT.\footnote{The code is available at {\color{blue}\url{https://github.com/d4v1d-cub/ApproxMasterEqKSAT.git}}}

\end{document}